\documentclass[twocolumn]{aa}
\usepackage{graphicx}
\usepackage{txfonts}
\usepackage{longtable}
\usepackage{natbib}
\begin{document}
   \title{A uniform X-ray analysis of 79 distant galaxy clusters \\
     with {\it ROSAT} and {\it ASCA}}

   \author{Naomi Ota \inst{1,2} \fnmsep\thanks{Present address: Cosmic Radiation Laboratoy, RIKEN,
       2-1 Hirosawa, Wako, Saitama 351-0198, Japan} 
       \and Kazuhisa Mitsuda\inst{2} }

   \titlerunning{X-ray analysis of 79 distant galaxy clusters}
   \authorrunning{N. Ota and K. Mitsuda}
   \offprints{N. Ota}

   \institute{Department of Physics, Tokyo Metropolitan University,
     1-1 Minami-osawa, Hachioji, Tokyo 192-0397, Japan\\
     \email{ota@crab.riken.jp}
     \and
     Institute of Space and Astronautical Science, Japan Aerospace
     Exploration Agency, 3-1-1 Yoshinodai, Sagamihara, Kanagawa 229-8510, Japan\\
     \email{mitsuda@astro.isas.jaxa.jp}
   }

   \date{To appear in A\&A}

   \abstract{ We present a uniform analysis of the {\it ROSAT} HRI and
     the {\it ASCA} GIS/SIS data for 79 distant clusters of galaxies
     in the redshift range $0.1 < z < 0.82$ to study the global
     structures of the intracluster medium. We have constructed an
     X-ray catalog consisting of the largest sample of clusters in the
     redshift range for which pointed X-ray observations were carried
     out with both the observatories. We determined the
     emission-weighted X-ray temperatures of the clusters with {\it
       ASCA}, while we studied surface brightness distribution with
     the {\it ROSAT} HRI utilizing the isothermal $\beta$ model.  We
     investigated the statistical properties and trends for redshift
     evolution of the X-ray parameters including the temperature, the
     density profile of the intracluster gas and the gas-mass fraction
     within $r_{500}$.  We also present correlations of the
     cluster parameters with the X-ray temperature and with the core
     radius and compare them with the predictions of the self-similar
     model, from which we discuss the possible origin of the double
     structure discovered in the core radius distribution.

     }

   \maketitle

\section{Introduction}

Clusters of galaxies are the largest collapsed systems known in the
universe. Because the time scale of evolution of clusters is a
significant fraction of the age of the Universe, the clusters may
preserve information about the early universe and thus are considered
to be excellent tracers of the formation and evolution of
structures. They are considered to continue to grow into larger
systems through complex interaction between smaller systems, namely
merging process. It is possible that the clusters that we see 
are in different stages of evolution.

For the understanding of cluster structure and evolution, we believe
it is important to analyze systematically a large number of 
clusters at various redshifts.  After the {\it ASCA}
\citep{Tanaka_etal_1994} and the {\it ROSAT} \citep{Truemper_1993}
X-ray observatories were put into orbit, it became possible to study
relatively distant clusters at X-ray energies. During 7--10 years of
observations, more than one hundred clusters were recorded
with both observatories.  Since {\it ASCA} has a high sensitivity to
measure the X-ray spectrum in the wide energy band while {\it ROSAT}
is good at imaging in the soft X-ray band, the two observatories are 
an excellent combination to study properties of the intracluster medium 
(ICM).  At present the {\it XMM-Newton} and the {\it Chandra}
satellites are in orbit and generate much cluster data with higher
sensitivities. However, as we will mention below, the data set used in
the present paper will be one of the best existing to construct
the largest sample of distant clusters and study global X-ray
structures.

X-ray observations bring us valuable information on not only the physical
state of the ICM but also the underlying potential structure of the
clusters. Statistical studies are very powerful in exposing
the physical nature of the clusters.  In particular, nearby clusters
have been extensively studied at X-ray wavelengths.  As to the low
redshift samples ($z\lesssim0.1$), \cite{Mohr_etal_1999} performed a
systematic analysis on the {\it ROSAT} PSPC data of 45 clusters and
utilizing the published ICM temperatures, they investigated the
correlation between the ICM mass and the temperature, namely the
$M_{\rm gas}-T$ relation.  They found that the slope of the relation is
significantly steeper than that predicted from the self-similar model
\citep{Kaiser_1986}. The observed luminosity-temperature ($L_X-T$)
relation is also known to be steeper than the expectation of the
self-similar model \citep[e.g.][]{David_etal_1993}. The inconsistency
between the observations and the simple theoretical model has been
debated for many years and various possibilities such as
non-gravitational heating \citep[e.g.][]{Evrard_Henry_1991,
  Cavaliere_etal_1997} and dependence of gas mass or gas-mass fraction
on the temperature have been proposed 
\citep[e.g.][]{David_etal_1993,Neumann_Arnaud_2001}.  For higher
redshift samples, however, the number of clusters that were uniformly
analyzed was limited compared to the nearby clusters. The {\it ASCA}
spectroscopic data of distant clusters were compiled by
\cite{Mushotzky_Scharf_1997} (38 clusters with $z>0.14$),
\cite{White_2000} (41 clusters with $z>0.1$ and 65 clusters with
$z<0.1$), and \cite{Novicki_etal_2002} (32 clusters with $0.3<z<0.6$
  and 53 clusters with $z<0.3$), while the combined (i.e. spectral and
  imaging) analyses of the distant clusters were separately published
  in \cite{Allen_1998} (13 clusters with $0.1<z<0.45$),
  \cite{Hashimotodani_1999} (27 clusters with $0.1<z<0.78$),
  \cite{Ettori_Fabian_1999} (36 clusters at $z>0.05$),
  \cite{Schindler_1999} (11 clusters at $z>0.3$),
  \cite{Lewis_etal_1999} (14 clusters with $0.14<z<0.55$) and 
  \cite{Vikhlinin_etal_2002} (22 clusters at $z>0.4$).  Recently,
  \cite{Ettori_etal_2004} reported the {\it Chandra} analysis of 28
  clusters at $0.4<z<1.3$ and observed the steeper slopes in the $L-T$
  and the $M_{\rm gas}-T$ relations, which provided hints of
  negative evolution in their relations at high redshift.

On the other hand, \cite{Ota_2001} and \cite{Ota_Mitsuda_2002} have
attempted to construct the largest X-ray sample of distant clusters
with $z>0.1$ based on the combined analysis of the {\it ROSAT} HRI and
the {\it ASCA} GIS/SIS data. The data set used in the analysis is one
of the best suited for the purpose of the present systematic study for
the following reasons: the largest cluster sample of the pointed
observations were stored in the {\it ROSAT} and the {\it ASCA}
archival database, which enable us to cover the widest redshift range
up to $z\sim 1$.  The focal plane instruments, the {\it ROSAT} HRI and
the {\it ASCA} GIS/SIS have sufficient sensitivities to study the
global properties of the ICM spatial structure with a typical
resolution of $~5\arcsec$ and spectral features such as the ICM
temperature and the luminosity, respectively.  In
addition, the instruments' responses were well calibrated and the use
of the same detectors for all the clusters will reduce systematic
effects when comparing their results. Our sample covers the
intermediate redshift range of 0.1--0.8 between those of Mohr et al.'s
sample ($z<0.1$) and Ettori et al.'s sample ($0.4<z<1.3$), thus
combining our data with the other two samples is 
helpful in discussing the
evolution of internal structure of the ICM directly from the
observational point of views.

In this paper we describe a uniform analysis of the {\it ROSAT} HRI and
the {\it ASCA} SIS/GIS data of 79 clusters performed in
\cite{Ota_2001} and \cite{Ota_Mitsuda_2002} and thus provide
an X-ray database of ICM structure with the widest
redshift range of $0.1<z<0.82$. In order to perform a reliable
parameter determination we paid special attention to evaluating all
possible systematic errors in the spatial and the spectral
analyses. We determined the spectral and spatial properties of the
intracluster gas in terms of the temperature, the core radius and the
central electron density etc. for all of the clusters. In the image
analysis we found that the sample can be naturally classified into
regular and irregular clusters according to the X-ray surface
brightness distribution. We present the statistical properties of the
X-ray parameters and the relations to the double-$\beta$ nature of the
clusters discovered in the core radius distribution. We also estimated
the cluster total mass, the gas mass and the gas-mass fraction within
a radius corresponding to a fixed overdensity of 500.  We further
studied the redshift dependency of the parameters and their
correlations, where we considered a systematic error due mainly to the
limited sensitivity of the instruments and some assumptions used in
the estimations.

This paper is organized in the following manner. In the next section,
we describe the characteristics of our samples. In section
\ref{sec:spatial_analysis} and \ref{sec:spec}, we explain the methods
of spatial analysis with {\it ROSAT} and spectral analysis with {\it
  ASCA} in detail. In section \ref{sec:results}, we derive the X-ray
parameters to describe the global structure of clusters and
investigate their redshift dependence. In section \ref{sec:corr} we
study correlations of the parameters and discuss the properties of the
ICM.  In section \ref{sec:summary} we summarize our
results.

We use $\Omega_{\rm M} = 0.3$, $\Omega_{\Lambda}=0.7$ and
$h_{70}\equiv H_0/(70~{\rm km\,s^{-1}Mpc^{-1}})=1$. The quoted errors are the
90\% confidence range throughout the paper except where noted.

\section{The sample}\label{sec:sample}

We have selected distant clusters with $0.1\leq z \leq 1$ for which
pointed X-ray observation data with both {\it ASCA} and {\it
  ROSAT}/HRI are available.  Though there are 83 clusters which meet
the criteria, we rejected three (\object{A222},
\object{A223},\object{A1758S}) because of the large ($>210''$)
pointing off-axis angles in the {\it ROSAT} observations. We did not
include a high-redshift cluster, \object{AXJ2016+112} at $z=1$
\citep{Hattori_etal_1997} because the {\it Chandra} observation showed
that the X-ray emission from the direction of AXJ2016+112 can be
resolved into point sources and the cluster diffuse emission is not
prominent \citep{Chartas_etal_2001}.  The final sample comprises 79
clusters. Among them, three clusters (\#1 \object{PKS0745-19}, \#6
\object{A2204}, and \#13 \object{A1689}) overlap with
\cite{Mohr_etal_1999}'s nearby cluster sample and thirty are known to
have strong gravitational lensing signals
\cite[e.g.][]{Hattori_etal_1999}. The observation logs are summarized
in Table 1, 
where the clusters were sorted according to the
redshift and numbered from 1 to 79. The redshift distribution of the
sample is shown in Fig. \ref{fig1}a. The clusters with $0.1\leq
z<0.3$, $0.3 \leq z<0.5$ and $0.5 \leq z < 1$ make up 58\%, 32\%, and
10\% of the sample, respectively. The average redshift is 0.30.

\begin{figure} 
\centering
\caption{(a) Redshift distribution of 79 distant clusters. 
(b) Redshift distribution of the 45 regular clusters (open) 
and the 34 irregular clusters (hatched).
\label{fig1}} 
\end{figure}

Because our analysis targets were collected from the proposal
observations and the sensitivities for high-redshift clusters are
limited, we have to carefully consider possible selection bias. In the
first step, we compare the sample with other unbiased cluster samples.
We show temperature distributions of our sample and the nearby X-ray
flux-limited 55 cluster sample constructed by \cite{Edge_etal_1990} in
Fig. \ref{fig2}. Our sample covers the equivalent temperature range,
but has a higher average temperature of 6.8 keV. The
Kolmogorov-Smirnov (K-S) test gave the probability that the two
samples are from the same temperature distribution as 0.06 (the K-S
parameter, $D = 0.24$).  Observation bias will be discussed in
\S\ref{subsubsec:param_spec} and \ref{subsubsec:param_beta} in more
detail.

\begin{figure} 
\centering
\caption{Comparison of the temperature distribution of our 79 distant
clusters (open) to the nearby flux limited 55 sample of
\cite{Edge_etal_1990} (hatched).
\label{fig2}} 
\end{figure}

\section{Spatial analysis}\label{sec:spatial_analysis}

\subsection{Data reduction from event lists}

We have retrieved the event lists of the {\it ROSAT} HRI detector from
the {\it ROSAT} Data Archive of the Max-Plank-Institut f\"{u}r
extraterrestrische Physik (MPE) at Garching, Germany.  We used the
EXSAS analysis package \citep{Zimmermann_etal_1992} to produce X-ray
images from the event lists. The raw HRI data has a 0\arcsec.5 spatial
resolution.  However, since the half power diameter of the X-ray
telescope is 4\arcsec.8 at the optical axis and the photon counting
statistics of the present data are limited, it is not worth
oversampling the telescope point spread function (PSF). We thus
rebinned the image into $5\arcsec$ bins, within whose diameter about
70\% of photons from a point source are included.

If there were multiple pointings for a cluster, and the pointing
offset angles between them are smaller than $3\arcmin$, we superpose the
event lists in the sky coordinate. The PHA channel of 1--10, which
corresponds to 0.2--2 keV, was used to avoid particle background
events.

Next we searched for foreground/background sources in the field of
view with the standard source detection program in the EXSAS. We
created lists of all the sources detected by the maximum likelihood
technique and having a likelihood of $>8$. We used the lists 
to exclude those sources from the analysis.

\subsection{Centroid determination and X-ray morphology}

We apply a method to determine cluster centroids and classify the
morphology of clusters.  A similar technique to find the cluster
center was first introduced by \cite{Mohr_etal_1993}.  We extended the
method to evaluate the systematic errors of the centroid
determination for distant clusters and found that the analysis can
also be used to determine the regularity of the X-ray distribution.  The
cluster emission is typically extended about $4\arcmin$ from the
optical axis of the telescope. In this image region, the position
dependence of the telescope vignetting is known to be less than 2\%
\citep{Briel_etal_1997}. The position dependence of the background
intensity is also small up to $\sim12\arcmin$. Thus they do not affect
the centroid determination.

The analysis consists of two major steps. First we estimate the 0-th
order center position and a parameter which represents the extent
of the X-ray image, utilizing 1-dimensional image projections and
Gaussian fits to them.  Then, starting from the 0-th order initial
value, we determine the center from the center of gravity of the
photon distribution.

First, we extracted a $4\arcmin\times4\arcmin$ image that contains the
central region of the cluster emission. Then we projected along the $x$-
and $y$- axes to get one-dimensional intensity profiles. We fit these
with Gaussian functions, and determined the mean, $(x_{G,0}, y_{G,0})$
and the width, $(\sigma_{x,0},\sigma_{y,0})$. To determine
these values with higher accuracy, we extracted an image of size
$3\sigma_{x,0}\times 3\sigma_{y,0}$ whose center is at $(x_{G,0},
y_{G,0})$ and performed the one-dimensional Gaussian fitting again to
derive the next set of $(x_{G,1}, y_{G,1})$ and
$(\sigma_{x,1},\sigma_{y,1})$.  The procedure was iterated $i$ times
until the mean position converged within 0.1 pixels i.e. $|(x_{G,i},
y_{G,i}) - (x_{G,i-1}, y_{G,i-1})| < (0.1, 0.1)$ or the number of
iterations reached $i=20$. We then define a parameter $\bar{\sigma}$ as
$\bar{\sigma} \equiv (\sigma_{x,i} + \sigma_{y,i})/2$, which is a
measure of the image extent for further analysis.

Next we determine the cluster center from the center of gravity of the
photon distribution in an aperture circle of radius, $R$. Then if
the X-ray image is circularly symmetric and the center of the circle is
at the X-ray center, the center of gravity of the photon distribution
should coincide with the center of the circle.  For a given value of
$R$, we can determine the center position, $\vec{r_{i}} \equiv
\sum_{R} \vec{p}/\sum_{R} 1$, where $\vec{p}$ represents the position
of the photon, and $\sum_{R}$ sums all the photons over the circular
area of the radius $R$. Then starting with the mean position
determined in the previous paragraph, $\vec{r_0}=(x_{G,i}, y_{G,i})$,
we extract a circular image of the radius $R$ centered at $\vec{r_0}$,
and calculate the centroid position $\vec{r_1}$.  We continue the
iteration until $|\vec{r_{i}}-\vec{r_{i-1}}|$ becomes less than 0.1
pixels.  If there were contaminating sources in the circle, we
excluded the region centered at the sources and the region symmetric
to them so as not to affect the centroid determination.  We changed
the aperture radius $R$ from 2 $\bar{\sigma}$ to 9 $\bar{\sigma}$ to
study the $R$ dependence of the results. As a result, while some
clusters showed constant centroids almost independent of the radius,
others exhibited systematic behavior. Two representative cases are
shown in Fig. \ref{fig3}.

\begin{figure*} 
\centering
\caption{Centroid determination and classification of X-ray
morphology. \#13 \object{A1689} and \#19 \object{A2163} are
shown in the left and right panels, respectively. From top to bottom:
(a) the HRI image in the pixel coordinates.  (b) The centroids
determined in circles of radii $R=n\bar{\sigma}$ ($n=2,3,..,9$), where
the size of the symbols are nearly proportional to $n$.  (c) The
deviations the centroids relative to that determined for
$3\bar{\sigma}$ in units of standard error, shown as a function of 
$R~[\bar{\sigma}]$. While \object{A1689} is classified as a 
regular cluster, \object{A2163} is an irregular cluster due to
the significant centroid deviation. \label{fig3}} 
\end{figure*}

The dependence of the center of gravity on the aperture radius may
indicate some asymmetry or substructures of the cluster.  However,
because the results with different $R$ are not statistically
independent, the Poisson fluctuations may produce systematic
deviations.  To estimate this effect, we performed
Monte-Carlo simulations and generated a number of simulation images
assuming the isothermal $\beta$-model as the X-ray image distribution.
We calculated the RMS value of the distance between the center of
gravity and the true center, $\sigma_{\rm MC}$, for various
combinations of parameters, the cluster core radius $r_c$, the number
of photons $N$ and the size of the aperture $R$, while $\beta$ was
fixed to the typical value of 0.67.

If the displacement of the center of gravity with different values of the
aperture radius $R$ is larger than the standard deviation determined
from the simulation, we can conclude that there is asymmetry or
substructures.  In Fig. \ref{fig3}, we show the deviation of the
center of gravity from that determined for $R=3\bar{\sigma}$
normalized by the standard deviation.  For some clusters the deviation
is well within the $1\sigma_{\rm MC}$ level; however, for others it is not.
We define criteria for irregularity of clusters as follows: (1) more
than 1 data point whose deviation exceeds $3\sigma_{\rm MC}$, and/or
(2) deviations ($>1\sigma_{\rm MC}$) seen in more than a few
consecutive points. Accordingly, we classified our sample into regular
and irregular clusters. The results are listed in Table 2. 
The ratio of regular to irregular is 45 : 34 and
the redshift distributions of the two subgroups are shown in Fig.
\ref{fig1}b.

\subsection{Radial X-ray surface brightness profiles}

We derive azimuthally-averaged radial profiles of the X-ray surface
brightness centered at the cluster centroids that were determined
within the $3\bar{\sigma}$ aperture radii for both the regular and
irregular clusters. We have chosen the centroids because they are less
affected by the Poisson fluctuations in the outer image regions. The
bin size of the radial profile is 5\arcsec. We excluded the
contaminating sources from the integration area with a circle of radius
5 times the FWHM of the PSF at position.

We study the effect of the choice of the center on the results of
the radial profile fits, by shifting the center positions determined
for $R$ between $2\bar{\sigma}$ and $9\bar{\sigma}$. We found the
$\beta$-model parameters (\S \ref{subsec:sbetafit}) constant
within the statistical errors for all the regular clusters and many of
the irregular clusters. For 30\% of the irregular clusters, the resulting core
radius increases typically by a factor of 2 as $R$ increases from
$2\bar{\sigma}$ to $9\bar{\sigma}$.

\subsection{Radial profile fitting with a single $\beta$-model}
\label{subsec:sbetafit}

In the following two subsections, we analyze the X-ray radial profile
under the isothermal $\beta$-model \citep{Cavaliere_Fusco_1976}. The
single $\beta$-model fitting function is written as
\begin{equation}
S(r) = S_0\left[ 1 + \left(\frac{r}{r_c}\right)^2\right]^{-3\beta+1/2} + C ,
\label{eq:single_beta}
\end{equation}
where $S_0$, $r_c$ and $\beta$ are the central surface brightness,
core radius and the outer slope, respectively, and $C$ is a constant
background. In Fig. \ref{fig4}a we show an example of
the radial profile fit.

\begin{figure} 
\centering
\caption{Radial X-ray surface brightness profile of \#13
\object{A1689} fitted with the single $\beta$-model (a) and the
double $\beta$-model (b). The crosses denote the observed radial
profile of the {\it ROSAT} HRI, and the step functions show the
best-fit $\beta$-models. The best-fit background levels are
shown with the horizontal dashed lines.  In the panel b), the
contribution of the inner and the outer component of the double
$\beta$-model are also shown with the dash-dot and the dotted lines,
respectively.\label{fig4}}
\end{figure}

It is crucial for reliable determination of the model parameters to
estimate the background level correctly. The HRI background is
dominated by the particle background \citep{Briel_etal_1997} and the
detailed calibration by \cite{Snowden_1998} showed that it can be
modeled by a constant image within $\sim12'$ from the detector
center. The counting rate of the particle background depends strongly
on the satellite orbit and time, and typically ranges from 1 to 10
${\rm counts\,s^{-1}}$ over the entire detector. Thus the background counting rate
needs to be determined observation by observation. We determined the
background level from the radial profile including the background as
one of the fitting parameters as Equation \ref{eq:single_beta}. Since
we do not know the true extent of the cluster emission and also $r_c$
and $\beta$ are strongly coupled (Fig. \ref{fig5}), the background
level and the model parameters depend on the outer radius of the
fitting area, $r_{\rm out}$. Particularly when $r_{\rm out}$ is too
small, the background level determined from the fit tends to be over-
or under-estimated and results in uncertain $r_c$ and
$\beta$. However, as shown in Fig. \ref{fig6}, the best-fit parameters
converge to constants if $r_{\rm out}$ is large enough, typically
$\gtrsim 10\bar{\sigma}$. For all the clusters, we confirmed that the
parameters converge at $r_{\rm out}=12\arcmin$. Thus we decided to
adopt this value for all clusters, within which the background can be
regarded as constant. Finally the background level we obtained from
fits are reasonable background levels.

\begin{figure} 
\centering
\caption{$\chi^2$ contour of the single $\beta$-model fit for \#13
  \object{A1689}.  The position of $\chi^2$ minimum is denoted with a
  cross. The curve corresponds to the single-parameter error domain at
  90 \% confidence.
\label{fig5}} 
\end{figure}

\begin{figure} 
\centering
\caption{Effect of the outer cut off radius in the single $\beta$
  model fitting. The results for \#13 \object{A1689} are shown as an
  example. The four parameters of the single $\beta$-model, $S_0~{\rm
    [counts\,s^{-1}arcmin^{-2}]}$, $\beta$, $r_c~{\rm [arcsec]}$, $C~{\rm
    [10^{-3}counts\,s^{-1}arcmin^{-2}]}$ are shown as a function of the outer
  cutoff radius, $r_{\rm out}$ in unit of $\bar{\sigma}$. The results
  of the constant fits to the data points at $r_{\rm out}\ge 10$ are
  shown with the dotted lines. \label{fig6}}
\end{figure}

Because the effective area of the X-ray telescope decreases with 
off-axis angle, the cluster image may be affected by the
vignetting. However since the off-axis angle of the cluster centroid
of the present data is smaller than $4\arcmin$, and the typical
spatial extent of the present clusters is $\sim 4\arcmin$, the
vignetting effect is at most 3 \% at the rim of the clusters (the HRI
vignetting function is given in \cite{Briel_etal_1997}). Although we
performed fits with the $\beta$-model fitting function with and
without the correction of the vignetting function, the results showed
no difference.

The $\beta$-model function needs to be convolved with the X-ray
telescope PSF, then integrated over the image bin.  However, since the
present image bin size is larger than the extent of the PSF, the
convolution with the PSF is not important.  Also, the $\beta$-model
function varies slowly within the $5\arcsec$ bins in most cases, so
integration within the image bin can be replaced by the value at the
center of the bin.  We confirmed these with simulations by comparing
two cases with and without convolution in the fitting model.  As
shown in Fig. \ref{fig7}, we find the difference of the two cases to be 
negligibly small as long as the core radius is larger than the bin
size, $5\arcsec$.  We also confirmed that both cases well
reproduce the assumed $\beta$ value of 0.6 within reasonable
statistical errors.  Thus in order to make the computation time
shorter, we skip the convolution with the PSF and the integration over
the bin. If the best-fit value (and the error domain) of the core
radius is smaller than the bin size, we regard it as an upper limit.

\begin{figure}
\centering
\caption{Reproducibility of $r_c$ in the single $\beta$-model
  analysis.  The $x$-axis is the assumed core radius in the simulation
  cluster image, and the $y$-axis is the core radius derived from
  fitting with two different models: the $\beta$-model with and
  without the PSF convolution. The results for three different photon
  counts are shown in each panel.}
\label{fig7}
\end{figure}

In Fig. \ref{fig5}, we showed a $\chi^2$ contour map on the
$\beta-r_c$ plane where the other free parameters, $S_0$ and $C$, are
optimized at each point of the plane.  The two parameters are strongly
coupled and the allowed parameter range exists in an elongated
region. We quote the 90\% confidence intervals on the best-fit
parameters. Table 2 
lists the results of single $\beta$-model
fitting. For \#43 \object{A1758} and \#79 \object{MS1054.5-0321},
because the fitting parameters did not converge properly, we optimized
the parameters within the range of $\beta \leq 3$. 

We compared the results of $\beta$-model fitting with
\cite{Hashimotodani_1999} for 27 {\it ROSAT}/HRI clusters with
$0.1<z<0.78$ and \cite{Ettori_etal_2004} for 10 high redshift {\it
  Chandra} clusters with constrained model parameters.  We found that
there is a good agreement for $\beta$ and $r_c$ measurements between
our and Hashimotodani samples within their statistical
errors. Furthermore for 8 of the 10 high redshift clusters, there is a
good agreement between our and the Ettori et al. samples within the 90\%
statistical errors.  For the two highest redshift samples, \#70
\object{RXJ1347.5-1145} ($z=0.451$) and \#71 \object{3C295}
($z=0.4641$), $r_c$ is systematically smaller by about 40\% compared
to the {\it Chandra} measurements. $\beta$ is also systematically
smaller with the mean ratio of 0.93 for 10 high redshift samples,
although the difference is the $\sim 2\sigma$ effect.  We consider that
the limited sensitivity of the HRI for the outer part of the cluster
emission may cause the underestimation of $\beta$ for the highest
redshift ($z>0.4$) samples.

As a result, the fractions of clusters with $\chi^2$ values exceeding
the 90\% and 99\% confidence levels are 25/79 (=0.32) and 11/79
(=0.14), respectively. The fractions are larger than expected only by
the statistical errors. Thus there may be some systematic errors that
are not well explained by the single $\beta$-model.  For some
clusters, systematic deviations from the current model are
particularly seen in the central regions (e.g. \#13 \object{A1689}),
which will be discussed in the next subsection.

We also evaluated an X-ray significance radius, $r_X$, representing
the extension of the observed cluster X-ray emission at which the
best-fit $\beta$-model surface brightness becomes equal to the
$3\sigma$ background level. The results are also listed in Table 5. 

\subsection{Radial profile fitting with a double $\beta$-model}
\label{subsec:wbetafit}

For some clusters, systematic residuals are seen in the results of
single $\beta$-model fitting.  As often seen in nearby clusters, this
may be attributed to the presence of central excess emission.  The
excess component is often represented by an additional $\beta$-model
component \citep{Jones_Forman_1984}, whose core radius is
$10\sim200~h_{50}^{-1}$kpc, and on average $60~h_{50}^{-1}$kpc
\citep{Mohr_etal_1999}. This corresponds to only several bins of the
present analysis even at $z=0.1$. It is not easy to constrain such
central emission in the distant clusters.  To evaluate
the statistical significance of the central emission, we attempted two
methods. We restrict this analysis to the 45 regular clusters because
the irregular clusters are often accompanied by substructures, which
can cause artificial double structures in their radial profiles.

In the first analysis, we exclude central bins from the fit and
investigate the variation of single $\beta$-model parameters against
the inner cutoff radius, $r_{\rm in}$. We test the improvement in the
$\chi^2$ value of the fit with the F-test compared to the case of
$r_{\rm in}=0$ (Fig. \ref{fig8}).  We define the value of $F$ as
$F=((\chi_1^2-\chi_2^2)/(\nu_1-\nu_2))/(\chi_2^2/\nu_2)$, where
$\chi^2$ and $\nu$ are the minimum $\chi^2$ value and the degrees of
freedom, and the suffix ``1'' and ``2'' correspond to the case of
$r_{\rm in}=0$ and $r_{\rm in}=n$ pixels, respectively ($\nu_1 - \nu_2
= n$).  For nine clusters, we find that the probability of exceeding
the $F$ value, $P$, rises at a certain 
$r_{\rm in}$ .  For seven such cases, the
core radius also starts increasing at the same inner cutoff
radius. This indicates that the nine clusters have significant two
core sizes.  Moreover for the seven of the nine the core radius of the
inner component at which the significant improvement of $\chi^2$
starts is roughly $r_{c}$.

\begin{figure} 
\centering
\caption{Effect of the inner cutoff radius in the single $\beta$ model
  fitting. The results for \#13 \object{A1689} are shown as an
  example. From top to bottom, the resultant values of $r_c$ [arcsec],
  $F$ (see \S\ref{subsec:wbetafit} for definition) and the probability
  of exceeding the $F$ value, $P$, are shown. \label{fig8}}
\end{figure}

In the second analysis, we assumed the double $\beta$-model composed of
different core radii ($r_{1} < r_{2}$),
\begin{equation}
S(r)=\sum_{i=1}^{2} S_i
\left[ 1 + \left(\frac{r}{r_{i}}\right)^2\right]^{-3\beta_i+1/2} + C ,  
\label{eq:fitmodel_db}
\end{equation}
and performed fitting to the radial profiles (Fig.
\ref{fig4}b). Because the inner slope $\beta_1$ is insensitive to the
fit, we linked it to the outer value, $\beta_1=\beta_2$. We tested the
statistical significance of the improvement of the $\chi^2$ value by the
F-test against the single $\beta$-model; $F = ( (\chi_{\rm
s}^2-\chi_{\rm d}^2) /(\nu_{\rm s}-\nu_{\rm d}))/(\chi_{\rm
d}^2/\nu_{\rm d})$, where the suffixes ``s'' and ``d'' stand for the
case of the single $\beta$-model and the double $\beta$-model,
respectively.  We found that nine of the regular clusters have significance
above the 95 \% level.  Those nine clusters are the same clusters for
which the double core nature is found in the previous analysis.  We
refer to the nine clusters with significant double structure as
``double-$\beta$'' clusters and show the result of the double
$\beta$-model fitting in Table 3.

We have plotted the two core radii against the core radius derived by
the single $\beta$-model for the nine ``double-$\beta$'' clusters in
Fig. \ref{fig9}a. It is remarkable that one of the two cores is
nearly consistent with the core of the single $\beta$-model, namely
$r_c\sim r_1$ or $r _c \sim r_2$. This indicates that the single
$\beta$-model tends to detect the one of the two components that is more
dominant. We also show the ratio of two cores versus the ratio of two
normalization factors in Fig. \ref{fig9}b. $r_2/r_1$ is $\sim 4$ on
average, while $S_2/S_1$ ranges from 0.01 to 1. These are 
consistent with nearby clusters \citep{Mohr_etal_1999}.  We can
classify them into two groups: inner core dominated clusters ($S_2/S_1
\lesssim 0.1$) and outer core dominated clusters ($S_2/S_1 \sim
1$). For the inner core dominant clusters, the single $\beta$-model
fit picks up the inner $\beta$-model component, while for the outer
core dominant clusters, the outer component is picked up. In Table 2 
the inner/outer core dominant clusters are denoted with W(1)/W(2).

\begin{figure}
\centering
\caption{Core radii derived by the double $\beta$-model for the nine
  clusters with significant double structure.  The panel a) shows
  the relation between two core radii of the double $\beta$-model and
  that of the single $\beta$-model, where the inner and outer cores of
  the double $\beta$ model are shown with the filled and open circles,
  respectively.  The panel b) shows the ratio of the two cores and
  the ratio of the two normalization factors in Equation
  \ref{eq:fitmodel_db}.  The two data points whose $S_2/S_1$ are large
  ($\sim 1$) in the panel b) correspond to the two clusters with
  large core radius of the single $\beta$-model in the panel a).
\label{fig9}} 
\end{figure}

We then investigated the reproducibility of the best-fit parameters of
the double $\beta$-model utilizing Monte-Carlo simulations.  We made
ten simulation clusters for each set of model parameters described
below and fitted them with the double $\beta$-model. We assumed $r_1 =
3~{\rm pixels}(=15\arcsec)$ as a typical core radius of the inner
component and several different values of $r_2$ between 6 and 24
pixels. We fixed $\beta_1 =\beta_2 = 0.67$.  For the intensity
ratio, we checked two typical cases, i.e. $S_2/S_1 =$ 0.1 and 1.0. We
then generated 75000 events within a radius of $12\arcmin$, which are 
typical total counts for double-$\beta$ clusters including
background.  We confirmed that the two core radii are well determined
as long as $r_2/r_1 \geq 3$. Thus,
at least for the nine double-$\beta$ clusters we detected, we can
conclude that the model parameters are well-determined by the model
fitting.

\subsection{Fraction of double-$\beta$ clusters}

We also find that the double-$\beta$ clusters are found only at
$z<0.3$ and the ratio to the regular clusters at $z<0.3$ is
32\%. However we have to take into account the fact that such double
structures are difficult to find at higher ($z>0.3$) redshifts because
of the low surface brightness.  In order to constrain the double
structure at higher redshifts, we estimated the upper limits of the
possible additional $\beta$-model component. For this purpose we
fitted the radial profile with double $\beta$-model with the ratio of
the two cores fixed at the average of nine double-$\beta$ clusters,
$r_{2}/r_{1}=4$.  We need to consider two cases: (1) the inner
component is dominant, namely the inner core radius is approximately
the core radius of the single $\beta$-model, $r_{1}\sim r_{c}$, and
(2) the outer component is dominant, $r_{2}\sim r_{c}$.  We thus
performed the fit with $r_{1}\sim r_{c}$ or $r_{2}\sim r_{c}$ as the
initial value.  For some cases, the fit converged to certain best-fit
values or upper limits of the fitting parameters.  However in some
cases the fit did not converge.  In such cases we fixed $r_{1}$ or
$r_{2}$ at the value of $r_c$ obtained from the single $\beta$-model
fitting.

We find that in addition to the nine double-$\beta$ clusters, there
are several other clusters that may contain a second
component.  In case (1) the outer component is marginally detected for
seven clusters at the 90\% confidence level.  For other 17 clusters,
the upper limit of the surface brightness of the outer component is
within the range $(0.01 - 0.1)S_1$ which is comparable to the range
for the seven inner-core dominant double-$\beta$ clusters. In case (2)
the inner component is marginally detected for three clusters, and
the upper limit is consistent with the two double-$\beta$ clusters for the
other 16. In Table 2
the clusters with marginal
inner/outer component are denoted with S(1)/S(2).

We finally obtain the fraction of clusters with marginal
double-$\beta$ structure at $z>0.3$ to be 35\%. It is comparable to
the occurrence of the double structures in the lower redshift systems
within the Poisson errors.  Thus we cannot conclude that there is
significant evolution of the fraction of the double-$\beta$ clusters
in the regular clusters within the observed redshift range.  On the
other hand, \cite{Ettori_etal_2004} noted based on the $\beta$-model
analysis that their high redshift ($z>0.4$) samples do not show any
significant double structure in the surface brightness
distribution. Thus there might be a trend of evolution in the core
structure of the ICM distribution starting around $z\sim 0.4$.  We
suggest that this should be confirmed by further observations.

\section{Spectral analysis}\label{sec:spec}

\subsection{Data reduction}
We retrieved the {\it ASCA} data sets from the High Energy
Astrophysics Science Archive Research Center (HEASARC) at NASA/Goddard
Space Flight Center 
and the DARTS Online Service
at the Institute of Space and Astronautical Science (ISAS)
that were screened with the standard REV-2
processing. We use the FTOOLS analysis package to reduce the cluster
spectra, and calculate the response functions of the telescopes and
the detectors.

The GIS was operated in the PH-nominal mode during observations. The
SIS has several choices between the CCD data modes (FAINT or BRIGHT)
and the CCD clocking modes (1CCD or 2CCD or 4CCD). For observations
done in a mixture of the FAINT and BRIGHT modes, we can combine
converted BRIGHT (on-board FAINT) and on-board BRIGHT mode data.

First we extacted the X-ray images in the 0.7 -- 10 keV for the GIS and
the 0.5 -- 10 keV for the SIS, respectively. The two sensors in the
same system, namely GIS-2, -3 and SIS-0, -1 are added together. We
fitted the projected images to Gaussian functions and determined the
peak positions. We accumulate spectra from a circular region centered
on the Gaussian peak, where the extraction radii are $6\arcmin$ and
$3\arcmin$ for the GIS and the SIS, respectively. We will discuss the
effect of the contamination from foreground/background point sources
in the next subsection.  We select a larger integration area for the
GIS because the FWHM of the point spread function of the GIS detector
alone depends on the incident photon energy $E$ keV, and is given by
$0.5\times(5.9/E)$ \citep{ABC_GUIDE}.  The FWHM for soft photons is
wider than for hard photons; the smaller integration radius would make
the spectrum harder, resulting in a systematically higher temperature.
This is a serious effect for a distant cluster with apparent core size
larger than $1\arcmin$ when the GIS integration radius is smaller than
$\sim 4\arcmin$.  Note that $6\arcmin$ corresponds to $\sim 1$ Mpc at
$z=0.1$, and it mostly covers the cluster region corresponding to
  $\Delta_c=500$ (See \S \ref{subsubsec:defs} for definition). The
systematic error in estimating the bolometric luminosity due to the
fixed integration radii will be discussed in the next subsection.

We subtract background spectra that were obtained during blank-sky
observations. Because the {\it ASCA} background has a detector position
dependency, we extract them from the same region as the cluster in the
detector coordinates.

The instrument response can be split into two parts: a
redistribution matrix (RMF), which specifies the channel probability
distribution for a photon of given energy, and an effective area curve
(ARF), which specifies the telescope area and window absorption.  We
utilized the latest version of the GIS RMFs, gis2v4\_0.rmf and
gis3v4\_0.rmf, while we generate the SIS RMF using the FTOOL
sisrmg. We built the ARF files with the ASCAARF program appropriate
for the cluster extended emission, summing the ARFs for each bin in
the cluster image region according to the weight of the photon counts.

\subsection{Fitting with Raymond-Smith model}

In order to measure the average, emission-weighted X-ray temperature
of the gas, we fitted the SIS and GIS spectra simultaneously with a
thin-thermal plasma emission model from \cite{Raymond_etal_1977}.
There are four parameters in the spectral model, the temperature $kT$,
the metallicity relative to the solar abundance $Z$, the redshift $z$,
and the normalization factor.  The redshift of each object was fixed
at the cataloged value in the NASA/IPAC Extragalactic Database (NED).
The fitting function was convolved with the telescope and detector
response functions.  In the spectral fitting, we used the XSPEC
version 9.0 analysis software \citep{Arnaud_1996}.  We rebinned the
spectral channels so that each bin contains at least 40 photons.

\cite{Yaqoob_1999} pointed out that fitting with fixed $N_{\rm H}$
result in a systematically high temperature because of the serious
decrease of the low-energy efficiency of the SIS since early 1994. To
avoid the problem, the absorption column density $N_{\rm H}$ was
allowed to vary.  Note that in the observation of \#57 \object{A402},
which was done during the AO7 phase, a serious reduction in the
efficiency below 1 keV was seen in the SIS spectra, so we used only
the GIS data for this cluster.

We checked the contribution of foreground/background contaminating
sources in the {\it ASCA} spectra. We picked up the point sources
whose photon counts are greater than 10\% of the cluster from the HRI
source lists. Excluding regions of $r=1\arcmin$ circles around the
sources, we recalculate the spectra, responses and backgrounds to
determine the temperature by the Raymond-Smith model fitting. Note
that Energy Encircled Function at $1\arcmin$ is approximately $\sim
0.3$ for the GIS and $\sim 0.4$ for the SIS, respectively. We
estimated the difference of the best-fit temperatures with and without
point sources excluded relative to the $1\sigma$ error of the
measurement, and found that the contribution of the point-like sources
does not affect the temperature measurement for most cases, except
for \#56 \object{CL0500-24}, \#66 \object{CL0024+17}, \#76 \object{3C220.1}, and
\#20 \object{A963}. For \object{A963}, we excluded one nearby source from
the spectral region of the GIS and the SIS.  More specific analyses
are described in \cite{Ota_etal_1998} for \object{CL0500-24},
\cite{Soucail_etal_2000} for \object{CL0024+17}, \cite{Ota_etal_2000}
for \object{3C220.1}.  In the case of \object{CL0500-24} and
\object{CL0024+17}, only the SIS data were used. Table 4
lists the results of the spectral analysis.  

We compared our results with
values published in \cite{White_2000} (there are 26 clusters in common)
to find a good agreement between the two measurements with a mean
temperature ratio of 1.04. We also compared 6 high redshift clusters
with constrained {\it ASCA} temperature with \cite{Ettori_etal_2004}.
The temperatures for 5 clusters are statistically consistent between
the two results, while there is systematic difference for \#71
\object{3C295}. As for \object{3C295}, \cite{Ettori_etal_2004}
excluded the central emission associated with the AGN, thus our
temperature measurement may be affected by the AGN. 

We estimated the X-ray luminosity in the 2 -- 10 keV band from the GIS
flux, $L_{\rm X} (6')$, and converted it into the bolometric
luminosity, $L_{\rm X,bol}(6')$, using the emissivity of the
Raymond-Smith plasma model. In order to check the systematic error of
the luminosity estimation, we calculated the bolometric luminosity by
integrating the $\beta$-model surface brightness distribution
determined with the {\it ROSAT} HRI within the same integration area,
to find that two estimations are consistent within about 15\%.  Then
we derived the bolometric luminosity within $r_{500}$ (see section
  \ref{subsubsec:defs}), $L_{\rm X,bol}$, by multiplying $L_{\rm
    X,bol}(6')$ with the ratio of the luminosities within $r_{500}$ to
  $6'$ using the $\beta$-model.  In Table 4 
we listed $L_{\rm X}(6')$, $L_{\rm X, bol}(6')$, and $L_{\rm X,bol}$.

\section{Cluster parameters}\label{sec:results}

In \S\ref{sec:spatial_analysis} we analyzed the cluster surface
brightness distribution under the assumption that the gas is
isothermal, and the spatial distribution is described by the single
$\beta$-model or the double $\beta$ model. In \S \ref{sec:spec}, we
determined the average X-ray temperature with the {\it ASCA} spectral
data. The X-ray images and the fitting results of individual clusters
are shown in Fig. \ref{figc1}.  
We will derive some physical
quantities of the clusters from the X-ray parameters obtained from the
analysis and investigate the redshift dependence of these parameters
and the histograms.

\subsection{Parameters from spectral analysis}
\subsubsection{Redshift dependence of spectral parameters}
\label{subsubsec:param_spec}

We show the spectral parameters obtained with the Raymond-Smith
fittings as a function of redshift in Fig. \ref{fig10}. In the plot
of the temperature, we quoted the results of nearby clusters
\citep{Mohr_etal_1999} for comparison. There is no significant change
in the temperature over a wide redshift range, particularly in
$z<0.5$. On the other hand, some clusters with $z>0.5$ resulted in
very high temperatures, though their errors are not well-constrained
due to the limited photon statistics and the error range overlaps with
the high temperature clusters in $z<0.5$.

\begin{figure*}
\centering
\caption{X-ray temperature(a), metal abundance(b), the 2--10 keV
  flux(c) and the 2--10 keV luminosity within $r<6\arcmin$(d)
  measured with {\it ASCA}. At $z>0.1$, the filled circles and the
  open triangles are the regular and the irregular clusters in our
  sample. The filled stars show the double-$\beta$ clusters in our
  sample. The {\it ASCA} sensitivity curve is indicated with the
  dotted line in the panel of $kT$, assuming the
  luminosity-temperature relation. The temperatures of clusters with
  $z<0.1$ were taken from \cite{Mohr_etal_1999}.  The asterisks and
  the filled squares denote the single-$\beta$ and double-$\beta$
  clusters in the nearby sample, respectively. }
\label{fig10}
\end{figure*}

As indicated from Fig. \ref{fig10}a, it is hard to detect a cluster
whose X-ray flux is below $\sim~10^{-13}~{\rm erg\,s^{-1}cm^{-2}}$ due to the
detection limit of {\it ASCA}. This flux corresponds to $\sim 1\times
10^{44}~{\rm erg\,s^{-1}}$ at $z=0.5$ and to $kT\sim2.5$ keV from the $L_X-T$
relation. Thus at $z>0.5$, clusters with temperature lower than 2.5
keV are barely detected. We consider that this can account for the
higher average temperature at $z>0.5$.

\subsubsection{Histograms of the spectral parameters}

In Fig. \ref{fig11} we plot the number of occurrence of each spectral
parameter to study how the samples are distributed in the parameter
space, regardless of the redshift. We show the distributions of the
regular and the irregular clusters separately as well as the
distributions of all the clusters. As a result, there is no clear
difference between the distributions of the regular and irregular
clusters. In Table 8 
we summarize the mean and the standard
deviation of the spectral parameters.

\begin{figure*}
\centering
\caption{Histograms of the spectral parameters determined with {\it
    ASCA} for 79 clusters (open). The panels (a) and (b) show the
  best-fit temperature and the metal abundance determined from the
  Raymond-Smith model fitting, respectively. The X-ray flux and the
  luminosity within $r<6\arcmin$ in the 2--10 keV band estimated
  with the GIS are shown in (c) and (d), respectively. The results of
  the double-$\beta$ clusters are superposed (hatched).}
\label{fig11}
\end{figure*}

\setcounter{table}{7}
\begin{table*}
\begin{center}
\caption{Means and standard deviations of cluster parameters\label{tab8}}
\begin{tabular}{lllllll}\hline\hline
Parameter & \multicolumn{2}{c}{Regular (45)} & \multicolumn{2}{c}{Irregular (34)} & \multicolumn{2}{c}{All (79)} \\ \cline{2-3} \cline{4-5} \cline{6-7}
 & Mean & Standard deviation & Mean & 
Standard deviation & Mean & Standard deviation \\ \hline
$kT$ [keV] & 6.3 & 2.8 & 7.2 & 2.7 & 6.8 & 2.8 \\
$Z$ [solar] & 0.33 & 0.22 & 0.24 & 0.15 & 0.29 & 0.19 \\
$L_{\rm X, bol}$ [erg/s] & $1.7\times10^{45}$ & $1.9\times10^{45}$ & $1.7\times10^{45}$ & $1.7\times10^{45}$ & $1.7\times10^{45}$ & $1.8\times10^{45}$ \\ 
$\beta$ & 0.56 & 0.11 & 0.73 & 0.61 & 0.64 & 0.32 \\
$r_c$ [$h_{70}^{-1}$ Mpc]  & 0.076 & 0.060 & 0.273 & 0.259 & 0.163 & 0.202 \\
$n_{e0}$ [$h_{70}^{1/2}\, {\rm cm^{-3}}$] & $3.6\times10^{-2}$ & $2.9\times10^{-2}$ & $7.4\times10^{-3}$ & $8.4\times10^{-3}$ & $2.4\times10^{-2}$ & $2.6\times10^{-2}$ \\
$\rho_0$ [$h_{70}^{2}\, {\rm g\,cm^{-3}}$] & $2.5\times10^{-24}$ & $2.5\times10^{-24}$ & $7.3\times10^{-25}$ & $2.7\times10^{-24}$ & $1.7\times10^{-24}$ & $2.7\times10^{-24}$\\
$t_{\rm cool}$ [Gyr] & 4.4 & 4.6 & 15.6 & 8.1 & 9.4 & 8.5 \\
$r_{500}$ [$h_{70}^{-1}$ Mpc] & 0.96 & 0.22 & 1.1 & 0.4 & 1.0 & 0.3 \\
$M_{500}$ [$h_{70}^{-1}\,{\rm M_{\sun}}$] & $4.3\times10^{14}$ & $5.1\times10^{14}$ & $9.7\times10^{14}$ & $2.4\times10^{15}$ & $6.7\times10^{14}$ & $1.6\times10^{15}$\\
$M_{\rm gas}$ [$h_{70}^{-5/2}\,{\rm M_{\sun}}$] & $7.1\times10^{13}$ & $4.3\times10^{13}$ & $9.5\times10^{13}$ & $6.5\times10^{13}$  & $8.2\times10^{13}$ & $5.5\times10^{13}$ \\
$f_{\rm gas}$ [$h_{70}^{-3/2}$] & 0.20 & 0.08 & 0.18 & 0.07 & 0.20 & 0.07 \\ \hline
\end{tabular}
\end{center}
\end{table*}

\subsection{Parameters from $\beta$-model analysis}
\subsubsection{Definitions of cluster parameters}\label{subsubsec:defs}

From the $\beta$-model fits and spectral analysis we obtained four
primary X-ray parameters $(kT, \beta, r_c, S_0)$ to describe the
intracluster gas.  From these parameters, we evaluate several
important quantities that characterize properties of the clusters and
the ICM. Below we summarize the definitions of these quantities.

\begin{enumerate}
\item Electron density at the cluster center : $n_{e0}$\\ The central
electron density can be determined from the central surface brightness
$S_0$, $\beta$, $r_c$ and the temperature.  However, from the present
analysis, we obtained the central surface photon flux convolved with
the telescope and the detector responses.  In this case $S_0$ can be
written as
\begin{eqnarray}
S_{\rm p0}(E_1,E_2) &=& \int_{E_1}^{E_2} dE' \int dE
 R(E',E)A(E) n_{e0} n_{\rm H0} \nonumber \\
 & & \cdot \frac{\Lambda_p(T,Z,E,z)\sqrt{\pi} r_c}{4\pi (D_A (1+z))^2}
\frac{\Gamma(3\beta-1/2)}{\Gamma(3\beta)},
\label{eq:beta_sp0_to_ne0}
\end{eqnarray}
where $S_{\rm p0}(E_1,E_2)~{\rm [counts~s^{-1}cm^{-2}]}$ 
is the photon flux in the $E_1 - E_2$
energy band, $R(E',E)$ and $A(E)$ respectively represent the response
function of the detector and the effective area of the X-ray
telescope/detector system, $\Lambda_p(T,Z,E,z)$ the emissivity in units
of ${\rm photons~s^{-1}cm^{3}keV^{-1}}$ for the object at redshift $z$. $D_A$
is the angular size distance to the cluster.  We utilized the
XSPEC program to perform the convolutions with the detector response
functions.  We adopt $n_{\rm H0}=(\mu_e/\mu_{\rm H})n_{e0}$, where
$\mu_{\rm H} = 1.40$ and $\mu_e = 1.167$.

\item Cluster mass profile and density profile: $M(r)$,
$\bar{\rho}(r)$, $\rho_{0}$\\ 
Assuming that the intracluster gas is in hydrostatic
equilibrium, the following condition is satisfied:
\begin{equation}
\frac{kT}{\mu m_p}\left( \frac{d\ln\rho_{\rm gas}}{d\ln r} 
+\frac{d\ln T}{d\ln r} \right) = -\frac{GM(r)}{r}. \label{eq:hydrostatic}
\end{equation}
In the case that the cluster gas is isothermal and has a $\beta$-model
density profile, $\rho_{\rm gas}(r)=\rho_{\rm
gas}(0)(1+(r/r_c)^2)^{-3\beta/2}$, 
where $\rho_{\rm gas}(0)=\mu_e m_p n_{e0}$, 
the total mass contained within the
radius $r$, is estimated from
\begin{equation}
M(r) = \frac{3kT\beta r}{\mu m_p
G}\frac{(r/r_c)^2}{1+(r/r_c)^2}.\label{eq:mvir}
\end{equation}
The average density within $r$ is then
\begin{equation}
\bar{\rho}(r) = \frac{M(r)}{\frac{4}{3}\pi r^3} = 
\frac{\rho_0}{1 + (r/r_c)^2}, \label{eq:average_rho}
\end{equation}
where $\rho_0 \equiv 9kT\beta/4\pi G\mu m_p r_c^2$ 
is the central total matter density.

\item Cooling timescale of the gas at the cluster center : $t_{\rm cool}$\\ 
We estimate the radiative cooling timescale of the
intracluster gas at the cluster center as
\begin{equation}
t_{\rm cool}=\frac{3k\sqrt{T}}{q_{\rm ff} n_{e0}}, \label{eq:tcool}
\end{equation}
where $q_{\rm ff}$ is related to the volume emissivity of thermal
Bremsstrahlung through $\epsilon_{\rm ff}=q_{\rm ff}n_e^2 T^{1/2}$.

\item Cluster limiting radius (overdensity radius) and cluster
  mass : $r_{500}$ and $M_{500}$ \\ We determine a cluster limiting
  radius within which the average density $\bar {\rho}(r)$ is equal to
  $\Delta_c$ times the critical density of the universe at the
  collapse time; namely
\begin{equation}\bar{\rho}(r) 
=\Delta_c \rho_{\rm crit}(z_{\rm col}). \label{eq:rvir}
\end{equation} 
We adopt a fixed overdensity of $\Delta_c =500$, which is
  justified in a sense that \cite{Evrard_etal_1996} suggested from
  their numerical simulations to use this value to study the gas
  properties and that the hydrostatic assumption is not valid beyond
  this radius, and that \cite{Finoguenov_etal_2001} showed that an
  assumption of isothermality also works at such overdensities.
Since we do not know the redshift of the cluster collapse, the most
simple assumption is that the clusters are observed just after they
are formed, i.e. $z_{\rm col} = z_{\rm obs}$. We will determine the
overdensity radius, $r_{500}$ under this assumption and calculate 
  the hydrostatic mass within $r_{500}$, $M_{500}$, from Equation
\ref{eq:mvir}.

\item Gas mass, and gas-mass fraction within $r_{500}$ : $M_{\rm
    gas}$ and $f_{\rm gas}$\\ The gas mass within $r_{500}$ is derived
  with
\begin{eqnarray}
M_{\rm gas} &=& \int^{r_{500}}_0 \rho_{\rm gas}(r) 4\pi r^2 dr \nonumber\\
&=& 4\pi \rho_{\rm gas}(0) {r_c}^3 
\int^{x_{500}}_0 (1+x^2)^{-3\beta/2}x^2 dx, 
\label{eq:mgas} 
\end{eqnarray}
where $x=r/r_c$ and $x_{500}= r_{500}/r_c$.
Then we obtain the gas-mass fraction with
$f_{\rm gas} = M_{\rm gas}/M_{500} $. 
\end{enumerate}

Among the four X-ray parameters, the temperature $kT$ is independently
determined from the other three $\beta$-model parameters. However,
the statistical errors of the three parameters are coupled to one
another.  In particular the coupling between $\beta$ and $r_c$ is
strong (Fig.  \ref{fig5}).  We determined the statistical errors of
the cluster parameters listed above with this coupling taken into
account.  For that purpose we first determined the error domain,
i.e. the statistically allowed parameter region, in the four
dimensional parameter space. Then, evaluating the cluster parameters
for all combinations of the X-ray parameters in the domain, we
determined the maximum and the minimum parameter values of the domain.
For the double-$\beta$ clusters, we also calculated those cluster
parameters from the double $\beta$-model. The methods of calculation
are similar to those shown above. The details are shown in Appendix
\ref{appendix:doubleb}.

\subsubsection{Redshift dependence of X-ray parameters}
\label{subsubsec:param_beta}

We plot the $\beta$-model parameters, and the parameters derived from
those X-ray parameters as functions of redshift in Fig.
\ref{fig12}.  We show the results of
double-$\beta$ model fits, and their inner and outer components are
distinguished by different symbols. In the figures we also plotted the
parameters taken from \cite{Mohr_etal_1999} for clusters with $z <
0.1$.

\begin{figure*}
\centering
\caption{Results from $\beta$-model analysis. In the panels (a)--(e),
$\beta$, $r_c$, $n_e$, $\rho_0$, and $t_{\rm cool}$ are shown.  At
$z>0.1$, the filled circles and the open triangles are the regular and
the irregular clusters in our sample. The filled stars and the open
stars show the inner core and the outer core of the double-$\beta$
clusters in our sample. At $z<0.1$ the filled squares
and the open squares denote the inner and outer components of the
double-$\beta$ clusters in the nearby \cite{Mohr_etal_1999} sample.
 The asterisks denote the nearby single-$\beta$
clusters. In the panel of $r_c$, the selection effects due to the
sensitivity and the spatial resolution of the {\it ROSAT} HRI are
indicated with the dotted line and the dashed line, respectively. In
the panel of $t_{\rm cool}$ we show a curve on which $t_{\rm cool}$ is
equal to the age of the Universe at the cluster redshift. }
\label{fig12}
\end{figure*}

\begin{figure*}
\centering
\caption{$r_{500}$, $M_{500}$, $M_{\rm gas}$, and $f_{\rm gas}$
derived from $\beta$-model analysis are shown in the panels
(a)--(d). See \S\ref{subsubsec:defs} for
definitions of the parameters. 
The meanings of the symbols are the same as in Fig.
\ref{fig12}.}
\label{fig13}
\end{figure*}

We do not see in those figures any clear redshift dependence in the
distributions of the X-ray parameters except for the parameters
related to $r_{500}$ shown in Fig. \ref{fig13}.  We will go
back to these parameters in section~\ref{subsubsec:rvir_fgas} 
and focus on the parameters that do
not involve the overdensity radius.

From Fig. \ref{fig12}b, we notice that the core radius shows an 
apparent redshift dependence.  As noted in \cite{Ota_Mitsuda_2002},
the core radius shows a remarkably large cluster-to-cluster
dispersion, spanning over two orders of magnitude. The core radii of
the irregular clusters are systematically larger than those of the
regular clusters, and there seems to be a gap in the $r_c$
distribution at around $0.1~h_{70}^{-1}$Mpc. 
The regular clusters also show a similar bimodal
distribution in $r_c$ but the fraction of the larger $r_c$ group
decreases with increasing $z$.

To investigate the selection effect, we created simulation
clusters with the Monte-Carlo method and performed the analysis on the
simulation clusters. We scale the count rate of
MS0906.5+1110 at $z=0.18$ to estimate the expected total
counts for a 40 ksec observation of a cluster at $z=0.5$ with a
typical luminosity $L_X\sim 1\times10^{45}~{\rm erg\,s^{-1}}$. The expected
cluster counts are about 600 counts. We simulated a series of cluster
images with various core sizes and found that the signal-to-noise
ratio is quite low for clusters with large cores of $r_c \gtrsim 400$
kpc at $z>0.5$. Based on the results we estimated the sensitivity of
the current HRI observation to be $S_0 \sim 3.6\times10^{-3}~{\rm
counts\,s^{-1}arcmin^{-2}}$. We show the sensitivity curve in Fig.
\ref{fig12}b. Thus the redshift dependencies of the core radius can be
explained by a selection effect.

Therefore we conclude that the X-ray parameters, temperature, core
radius, $\beta$, and the central electron density are consistent with
showing no significant trend of evolution at $z \lesssim 0.5$.

\subsubsection{Gas-mass fraction within $r_{500}$ 
and the systematic error}\label{subsubsec:rvir_fgas}

Although the X-ray parameters that are directly determined from 
observation do not show strong redshift-dependence, we find a weak
redshift dependence in the overdensity radius, which
  approximately follows $r_{500} \propto (1+z_{\rm obs})^{-0.6}$
(Fig. \ref{fig13}a).  The dependence is likely to be introduced when
we define $r_{500}$ by Equation \ref{eq:rvir}, i.e. $\bar{\rho}(r) =
\Delta_c \rho_c(z_{\rm obs})$. In other words, the redshift dependency
is introduced by the assumption, $z_{\rm col} = z_{\rm obs}$. 
We confirmed that such a negative dependence disappears when we
assume a constant $z_{\rm col}$ independently of $z_{\rm obs}$, for
example $z_{\rm col}=1$.

In Fig. \ref {fig13}d, we show the gas-mass fraction inside $r_{500}$
determined with $z_{\rm col}=z_{\rm obs}$. We do not see a clear
dependence on the redshift in our own data. We then obtain the average
gas-mass fraction of our sample of 79 distant clusters to be
\begin{equation}
\langle f_{\rm  gas}\rangle = (0.20\pm 0.07)~h_{70}^{-3/2}, \label{eq:fgas_r500}
\end{equation}
where the quoted error is the standard deviation of the
cluster-to-cluster variation.  On the other hand the average gas-mass
fraction of the nearby samples of \cite{Mohr_etal_1999} is $\langle
f_{\rm gas}\rangle_{\rm nearby} = 0.12 \pm 0.03$ (Note we recalculated
the value with Equation \ref{eq:rvir} and $\Delta_c=500$ under the
assumption of $z_{\rm col}=z_{\rm obs}$).  \cite{Mohr_etal_1999}
estimated their systematic errors to be $\sim 10$\%. Although our
distant sample seems to show higher $f_{\rm gas}$ values in comparison
to the nearby samples, the two results are consistent with each other
within their errors. If we further compare our result with the baryon
density in the Universe determined by the Wilkinson Microwave
Anisotropy Probe \citep{Spergel_etal_2003}, $\Omega_{\rm
  b}/\Omega_0=0.16$, it is again higher although they are in
  agreement within the errors. We will thus examine the possible
systematic errors of $f_{\rm gas}$ estimation due to (1) the
  choice of limiting radius,  (2) the calibrations of the X-ray
telescope/detector systems,  (3) the assumption of $z_{\rm col}=z_{\rm obs}$, and 
(4) the effect of the temperature gradient.

(1) We have defined the cluster limiting radius with  
  $\Delta_c=500$ and the resultant value is typically $r_{500}\sim
  1$~Mpc for the current sample.  On the other hand, the extent of the
  observed X-ray emission, $r_X$, is found to be larger for most of
  the clusters (Fig. \ref{figc1}) 
and $r_X/r_{500}= 1.5$ on average.
  Thus we do not need to worry about the effect of extrapolation in the
  current $f_{\rm gas}$ estimation.  The average gas-mass fraction
  within $r_X$ is derived as $(0.24\pm 0.06)~h_{70}^{-3/2}$, which
  agrees with Eq.~\ref{eq:fgas_r500} within the errors.

(2) Since the determination of the gas-mass fraction requires absolute
calibrations of the X-ray telescope/detector systems of {\it ROSAT}
and {\it ASCA}, we carefully examined the calibrational errors and  found 
that they can cause at maximum 25\% errors in $f_{\rm gas}$ (see
Appendix \ref{appendix:syserr} for details). 

(3) The assumption of $z_{\rm col}=z_{\rm obs}$ 
may be a source of uncertainty in $f_{\rm gas}$.
 We can infer $z_{\rm
  col}$ from the condition that the central mass density should be
higher than the average mass density, namely $\rho_0 >
  \bar{\rho}(r_{500})$. Since the observed range of $\rho_0$ is $\sim
2\times 10^{-26} - 1\times10^{-23}~{\rm g\,cm^{-3}}$, we obtain $z_{\rm
  col}\lesssim 1.3 - 18$, where the smaller (larger) value corresponds
to the clusters with large (small) core radii. Thus it is likely that
the clusters with large $r_c$ were formed at $z_{\rm col}\lesssim
1.3$.

We then vary the formation redshift, $z_{\rm col}$ for which 
we calculate the critical density, $\rho_{\rm crit}(z_{\rm col})$  
and find the radius where the measured matter density 
is 500 times $\rho_{\rm crit}(z_{\rm col})$.
If we simply assume a fixed formation redshift for all the clusters,
  ranging from 0.5 to 1.5, and calculate the mean gas-mass fractions
  within $r_{500}$ in the same manner as Eq.\ref{eq:fgas_r500},
  $\langle f_{\rm gas}\rangle$ varies from 0.18 to 0.12 with a typical
  standard error of 0.05.  
Thus the $f_{\rm gas}$ estimation 
  largely depends on the assumption of $z_{\rm col}$.
However such an effect is expected to be more serious for the low-redshift clusters, 
 whose average gas-mass fraction was measured to be $\langle
f_{\rm gas}\rangle_{\rm nearby} \sim 0.12$ 
\citep{Mohr_etal_1999, Sanderson_etal_2003}. 
Thus we may not attribute the systematic error in measuring 
$f_{\rm gas}$ for the distant sample 
to the assumption of $z_{\rm col}$. 

(4) The emission-weighted temperature reflects the temperature of the
cluster core region. Then, if there is a significant temperature drop
at the center, it may cause an overestimation of the gas-mass fraction
because the cluster hydrostatic mass estimation is more
sensitive to the temperature profile than the gas
mass. Such temperature drops were usually found in cluster cores with
short ($\sim$ a few Gyr) cooling timescales.  The spectral analysis of
the cooling flow clusters with the {\it XMM-Newton} and the {\it
  Chandra} satellites showed that the temperature drops typically by a
factor of 3 over the central $r\lesssim 200$ kpc region
\citep[e.g.][]{Tamura_etal_2001, Schmidt_etal_2001}.  We then
estimated the emission-weighted temperature within 1.5 Mpc
(corresponding to the typical integration radius for the GIS spectra)
assuming the radial temperature profile of $T(r)\propto r^{0.2}$ (from
Fig. 1 of \cite{Tamura_etal_2001}) and the $\beta$-model surface
brightness distribution with $r_c=50$ kpc and $\beta=2/3$ to find that
it is lower by about 30\% than that of the outer($r>0.1$ Mpc)
region. On the other hand, we obtained the mean gas-mass fraction for
the 26 regular clusters with $t_{\rm cool}\leq 3$ Gyr to be
  $\langle f_{\rm gas}\rangle=0.22 \pm 0.08$, which is larger by 20\%
than that of the rest of the sample. Thereby we consider that, for at
most 1/3 of the samples, $f_{\rm gas}$ may be overestimated to some
extent, depending on the degree of the temperature gradient. However,
considering the fact that the correlation between $f_{\rm gas}$ and
$t_{\rm cool}$ is weak as well as that the ranges of $f_{\rm gas}$ for
clusters with short and long cooling timescales are not very
different, the effect of the temperature gradient in estimating the
mean gas-mass fraction for all the samples is suggested to be not
large compared to the cluster-cluster variation. 
For more accurate measurements of $f_{\rm gas}$, 
we need to constrain the temperature profiles 
for the individual clusters and reduce the measurement uncertainties. 

\subsubsection{Histograms of the X-ray parameters}

Since we found no significant evolution in the X-ray parameters, we
will investigate the distribution of X-ray parameters obtained from
the $\beta$-model analysis without distinguishing the clusters by
redshift. In Fig. \ref{fig14}, we show the histograms of the X-ray
parameters.  We find that both $kT$ and $\beta$ are distributed in
ranges smaller than 1 order of magnitude, while $r_c$ and $n_{e0}$ are
distributed over almost two orders of magnitude.

\begin{figure*}
\centering
\caption{Histograms of the X-ray parameters determined with the
single $\beta$-model for 79 clusters (open). The results of the
double $\beta$-model fittings for the nine double-$\beta$ clusters are
superposed in the panels, where the hatched and filled regions show
contributions of the inner core and the outer core, respectively.}
\label{fig14}
\end{figure*}

\begin{figure*}
\centering
\caption{Histograms of $r_{500}$, $M_{500}$, $M_{\rm gas}$ and
$f_{\rm gas}$ derived with the single $\beta$ model for 79 clusters
(open). The results of double $\beta$ clusters are superposed
(hatched).}
\label{fig15}
\end{figure*}

While the mean value of $r_c$ for all the clusters is $\langle r_c
\rangle =0.163~h_{70}^{-1}$Mpc, if we treat the regular and the
irregular clusters separately, we obtain $\langle r_c
\rangle=0.076~h_{70}^{-1}$Mpc and $\langle r_c\rangle
=0.273~h_{70}^{-1}$Mpc for the regular and the irregular clusters (see
Table 8). 
Thus these are different by a factor of 3. We also
notice that the distributions of the irregular and regular clusters
are not separated. Instead, the distribution of the regular clusters
has a double-peaked structure, whose core radii corresponding to the
two peaks are $50~h_{70}^{-1}$kpc and $200~h_{70}^{-1}$kpc
respectively.  Thus there is about a factor of four difference. The
peak of the larger core radius coincides with that of the irregular
clusters. One may consider that the regular clusters with a large core
radius were classified as regular because the counting statistics were 
not very good. However, we find that the statistics of irregular and
regular clusters with a large core are not very different. Thus it
is difficult to explain this coincidence just by statistics. The core
radius distribution when the nearby
\cite{Mohr_etal_1999} samples and the our distant samples were added
together was shown in \cite{Ota_Mitsuda_2002}.
 
In Fig. \ref{fig14}b, we also show the distributions of the core
radii of the double-$\beta$ clusters. We notice that the double-peaked
structure of the core radius of the regular clusters seems related to
the double-$\beta$ structure because the core radius of the
smaller-core component is distributed around the lower peak of the
regular-cluster core radius distribution, while the larger-core is
around the higher peak.

In section \ref{subsec:wbetafit}, we have shown that about 20\% of
regular clusters have significant double-$\beta$ structures, and for
about 60\% of regular clusters, the existence of a similar structure
cannot be rejected.  We have also shown that there are inner-core
dominant and outer-core dominant cases and that a single $\beta$-model
fit picks up the core radius of the dominant core (Fig. \ref{fig9}).
Thus we consider that the correspondence between the core radius
distribution of all the regular clusters and the distribution of
double-$\beta$ clusters is not just a coincidence but that it is
related to the double core nature of the regular clusters.

The electron density, the dark matter density and the cooling time at
the cluster center show similar double-peaked distributions. On the
other hand, as shown in Fig. \ref{fig15}, $r_{500}$ and the
other three parameters evaluated within $r_{500}$ are
distributed in ranges smaller than 1 order of magnitude. We will
discuss the correlations between the parameters in the next section in
more detail. We list the mean and the standard deviation for all the
parameters in Table 8.

\section{Correlations between cluster parameters}\label{sec:corr}

We investigate correlations of various cluster parameters with the
X-ray temperature, $kT$ in \S\ref{subsec:corr_kt} and also with the
core radius, $r_c$ in \S\ref{subsec:corr_rc}, which showed the
distinct double-peaked distribution and thus may provide a clue to
understand the structures of clusters. Possible
systematic errors will be considered in each subsection.

 In the following analysis, we first calculate the Pearson's
 correlation coefficient, $R_{XY}=\sum_i (X_i - \bar{X})\sum_i (Y_i -
 \bar{Y})/\sqrt{\sum_i (X_i - \bar{X})^2}\sqrt{\sum_i (Y_i -
   \bar{Y})^2}$, to measure the strength of the correlation. We take
 the logarithm of the two measured parameters, $x$ and $y$, namely
 $X=\log x$ and $Y=\log y$, except for the case of the $Z-T$ and the
 $\beta-T$ relations. Then if $|R_{XY}| \ge 0.3$, we derive the
 best-fit relation between the two parameters assuming the power-law
 function. In order to take into account the statistical uncertainties
 of both the $x$ and $y$ axes, we performed the $\chi^2$ minimization
 in the linear ($Y=aX+b$) fit by defining $\chi^2\equiv \sum_{i} (Y_i-
 (a X_i + b))^2/(a\sigma_{X,i}^2+\sigma_{Y,i}^2)$, where
 $\sigma_{X,i}$ and $\sigma_{Y,i}$ are the $1\sigma$ errors for the
 parameter $X_i$ and $Y_i$, respectively. Because in all the cases
 below the fits were not statistically acceptable due to large
 scatters of the data at the 90\% confidence level, the error ranges of the
 coefficients $a$ and $b$ were estimated from the dispersions of the
 data points around the model functions rather than the photon
 statistics. We excluded two irregular clusters, \#43 \object{A1758}
 and \#79 \object{MS1054.5-0321}, in the analysis except for the $Z-T$
 and the $L_{\rm X, bol}-T$ relations because their $\beta$-model
 parameters were not well constrained (\S\ref{subsec:sbetafit}),
 although they are plotted in Fig. \ref{fig16} and \ref{fig18}.  We
 then compare some of the resulting relations with the predictions of
 the self-similar model.

\subsection{Correlations with the gas temperature}\label{subsec:corr_kt}

In Fig. \ref{fig16}a--h, we show eight parameters derived from the
spectral and the image analysis, the gas mass, and the gas-mass
fraction etc. as a function of X-ray temperature.  We mainly show the
relations derived for $\Delta_c=500$ below.
In Table 7, 
we show the cluster parameters of the individual clusters 
 for the overdensities of $\Delta_c=500\Omega^{0.427}$ 
and $18\pi^2\Omega^{0.427}$ \citep{Nakamura_Suto_1997} 
and in Table 9 
the scaling relations
with and without consideration of the cosmological factor, 
$E_z=H_z/H_0$ as noted by \cite{Ettori_etal_2004}.
 Since we did not find any strong redshift
evolution in the observed X-ray properties, we first derive the
parameter correlations regardless of their redshifts.  We will discuss
the correlations in the case where we take into account the sample
redshifts later.  Note that $kT$ and $r_c$ are in
units of [keV] and [$h_{70}^{-1}$Mpc], respectively. 

\begin{figure*}
\centering
\caption{Relations of the metallicity (a), the bolometric luminosity
  (b), $\beta$ (c), the central electron density (d), the
    overdensity radius (e), the cluster mass (f), the gas mass (g),
  and the gas-mass fraction (h) with the X-ray temperature. In all the
  panels the results of the single $\beta$-model fitting are plotted
  and the single-$\beta$ regular clusters, the single-$\beta$
  irregular clusters, and the double-$\beta$ clusters are denoted with
  the filled circles, the open triangles, and the filled stars
  respectively.  The horizontal and the vertical error bars are
  $1\sigma$.  In the panels (b) and (e)--(h), the best-fit
  power-laws {for the entire sample} are shown with the solid
  lines. In the panels (a), (c) and (h) the sample means are indicated
  with the dashed lines.  In panel (b), the best-fit $L-T$
    relations for two subgroups with $r_c<0.1$~Mpc and $r_c>0.1$~Mpc
    are indicated with the dashed and dash-dotted lines, respectively
    (see section~\ref{subsec:corr_kt} for details).  In panels
  (e)--(g), the slopes expected from the self-similar model are shown
  with the dotted lines.}
\label{fig16}
\end{figure*}
\begin{table*}
\begin{center}
\caption{Scaling relations for the distant clusters\label{tab9}}
\begin{tabular}{llll}\hline\hline
         & \multicolumn{3}{c}{$\Omega_{\rm M} = 0.3, \Omega_{\Lambda}=0.7$, 
         $z_{\rm col}=z_{\rm obs}$ }\\ \cline{2-4} 
Relation 
& $\Delta_c=500$ 
& $\Delta_c=500\Omega^{0.427}$ 
& $\Delta_c=18\pi^2\Omega^{0.427}$ \\ \hline
$L_{X,bol}-T$
               & $6.53^{+8.34}_{-3.82}\times10^{42} (kT)^{3.08^{+0.48}_{-0.45}}$ 
               & $7.94^{+8.92}_{-4.58}\times 10^{42}(kT)^{3.01^{+0.47}_{-0.42}}$ 
               & $1.07^{+1.26}_{-0.55}\times 10^{43}(kT)^{2.92^{+0.40}_{-0.43}}$ \\
$M-T$ 
               & $1.64^{+0.35}_{-0.26}\times10^{13} (kT)^{1.68^{+0.10}_{-0.11}}$
               & $2.07^{+0.35}_{-0.43}\times 10^{13}(kT)^{1.68^{+0.13}_{-0.08}}$ 
               & $3.31^{+0.73}_{-0.60}\times 10^{13}(kT)^{1.69^{+0.11}_{-0.11}}$ \\
$M_{\rm gas}-T$  
               & $2.88^{+0.93}_{-0.64}\times10^{12} (kT)^{1.85^{+0.14}_{-0.15}}$
               & $4.32^{+1.28}_{-0.89}\times 10^{12}(kT)^{1.80^{+0.13}_{-0.15}}$ 
               & $1.10^{+0.29}_{-0.25}\times 10^{13}(kT)^{1.67^{+0.15}_{-0.13}}$ \\
$f_{\rm gas}-T$ 
               & $0.28^{+0.09}_{-0.06}(kT)^{-0.08^{+0.14}_{-0.14}}$ 
               & $0.33^{+0.10}_{-0.08}(kT)^{-0.11^{+0.15}_{-0.15}}$
               & $0.47^{+0.16}_{-0.15}(kT)^{-0.19^{+0.20}_{-0.16}}$\\    \hline
$E_z^{-1}L_{X,bol}-T$ 
               & $5.50^{+7.92}_{-3.24}\times10^{42} (kT)^{3.12^{+0.51}_{-0.48}}$
               & $7.24^{+9.15}_{-4.04}\times 10^{42}(kT)^{3.00^{+0.45}_{-0.45}}$
               & $1.01^{+0.99}_{-0.55}\times 10^{43}(kT)^{2.89^{+0.43}_{-0.39}}$ \\
$E_zM-T$ 
               & $1.73^{+0.36}_{-0.32}\times10^{13} (kT)^{1.71^{+0.11}_{-0.10}}$
               & $2.11^{+0.45}_{-0.37}\times 10^{13}(kT)^{1.73^{+0.11}_{-0.11}}$
               & $3.51^{+0.75}_{-0.62}\times 10^{13}(kT)^{1.71^{+0.11}_{-0.10}}$ \\
$E_zM_{\rm gas}-T$ 
               & $3.05^{+0.94}_{-0.72}\times10^{12} (kT)^{1.88^{+0.15}_{-0.15}}$
               & $4.62^{+1.18}_{-1.04}\times 10^{12}(kT)^{1.82^{+0.15}_{-0.12}}$
               & $1.19^{+0.30}_{-0.28}\times 10^{13}(kT)^{1.68^{+0.15}_{-0.12}}$ \\ \hline
\end{tabular}
\end{center}
\end{table*}

\begin{itemize}

\item $Z-T$ \\ In Fig. \ref{fig16}(a), there seems to be a slight
  decline of the metallicity, $Z$, against the temperature, however, we
  found the correlation coefficient, $R_{XY}$ to be $-0.23$ for 78
  clusters, where \#66 \object{CL0024+17} was not used in the fit
  because $Z$ was so uncertain that the value was fixed at 0.3 solar
  in the spectral fitting \citep{Soucail_etal_2000}.  Thus there is
  no clear $Z-T$ correlation. This is consistent with the previous
  results on the metallicity measurements for 21 clusters reported by
  \cite{Mushotzky_Loewenstein_1997}.  We thus indicate the mean
  metallicity of our sample, $\langle Z\rangle=0.29$, with the dotted
  line in the figure (see also Table 8).

\item $L_{\rm X,bol}-T$ \\
We find a significant correlation between the bolometric luminosity 
and the temperature, and $R_{XY}$ is 0.59 for 77 samples.  Note that
\#56 \object{CL0500-24} and \#66 \object{CL0024+17} are not
included in order to avoid the systematic difference in the
integration radius for the spectra because the only SIS spectral data
were used for these two clusters due to the serious contamination from
the point sources in the vicinity of the clusters (see
\S\ref{sec:spec}).  We thus obtain the temperature-luminosity relation
for the 77 clusters to be
\begin{equation}
L_{\rm X,bol}~{\rm [erg\,s^{-1}]} = 
6.53^{+8.34}_{-3.82}\times10^{42} (kT)^{3.08^{+0.48}_{-0.45}}.
\label{eq:lbol_t}
\end{equation}
$\chi^2/{\rm d.o.f}=1703/75$. As a result, the slope of the
relation is consistent with those previously published for the nearby
clusters \citep[e.g.][]{Markevitch_1998, Arnaud_Evrard_1999}.
However, if compared to the relation for higher redshift clusters by
\cite{Ettori_etal_2004}, they showed a steeper slope of $3.72\pm0.47$
($1\sigma$ error).

The large error in the normalization factor in Equation
\ref{eq:lbol_t} is affected by the intrinsic scatter around the
best-fit relation.  We found from Fig. \ref{fig16}b that the
normalization factors of the $L_{\rm X, bol}-T$ relation are
significantly different between the regular and the irregular
clusters: for a fixed temperature, the regular clusters tend to have a
larger luminosity in comparison to the irregular clusters.  The
difference becomes more evident if we divide the sample into two
subgroups according to the core radius ranges of $r_c <
0.1~h_{70}^{-1}$ Mpc and $r_c > 0.1~h_{70}^{-1}$ Mpc \citep[see
  also][]{Ota_Mitsuda_2002}.  If we fit them separately, fixing
the slope at the best-fit value of 3.08, we obtain
\begin{eqnarray}
L_{\rm X,bol}~{\rm [erg\,s^{-1}]} &=& 
9.12^{+1.02}_{-0.92}\times10^{42} (kT)^{3.08}~{\rm for}~r_c \leq0.1, \\ 
L_{\rm X,bol}~{\rm [erg\,s^{-1}]} &=& 
3.98^{+0.22}_{-0.21}\times10^{42} (kT)^{3.08}~{\rm for}~r_c >0.1,
\end{eqnarray}
and $\chi^2/{\rm d.o.f}=810.9/38$ and 199.4/37, respectively. Thus
the the normalization factor for the small core clusters is larger
than that for the large core clusters at the $10\sigma$ significance level.

\item $\beta-T$ \\ There seems to be a trend of larger $\beta$
  values for higher temperatures, as previously noted by
  \cite{Schindler_1999}.  However $R_{XY}=0.27$ for 77 clusters, hence
  the correlation is not clear.

\item $n_{e0}-T$ \\ We see from Fig. \ref{fig16}d that there is a
significant scatter of the central electron density, $n_{e0}$.  We
obtained $R_{XY}=-0.28$ and thus the correlation is not clear from the
data.  On the other hand, we notice $n_{e0}$ is strongly correlated
with the parameter $r_c$, which will be discussed in the next
subsection in more detail.

\item $r_{500}-T$ \\ 
A strong correlation between the overdensity radius and the temperature
is found ($R_{XY}=0.82$), whose best-fit relation is:
\begin{equation}
r_{\rm 500}~[h_{70}^{-1}{\rm Mpc}]
= 0.38^{+0.03}_{-0.02}(kT)^{0.53^{+0.04}_{-0.04}}, \label{eq:rvir_t}
\end{equation}
and $\chi^2/{\rm d.o.f}$ is 151.2/75. The power-law slope of $0.53\pm0.04$ is
consistent with the value predicted from the self-similar model, 0.5,
within the error range.

\item $M_{500}-T$ \\ 
There is a tight correlation between the cluster mass and the
temperature ($R_{XY}=0.88$).  The best-fit relation is
\begin{equation}
M_{500}~[h_{70}^{-1}{\rm M_{\odot}}]
=1.64^{+0.35}_{-0.26}\times10^{13} (kT)^{1.68^{+0.10}_{-0.11}},
\end{equation}
and $\chi^2/{\rm d.o.f}$ is 206.5/75. The slope of
$1.68^{+0.10}_{-0.11}$ is slightly steeper than that expected from the
self-similar relation, i.e. $M_{500}\propto T^{1.5}$.  However, from
Fig.~\ref{fig16}f, we consider that the degree of the departure from
the self-similar relation is marginal if compared to the typical
size of the statistical error bars of the data points.

 We compare the $M_{500}-T$ relation with those derived for two
  samples of nearby clusters (Table 1 of \cite{Finoguenov_etal_2001}).
  Because they calculated it under a different cosmology: $\Omega_0=1$
  and $H_0=50~{\rm km\,s^{-1}Mpc^{-1}}$, we calculate the relation using the same set
  of cosmological parameters and obtain $M_{500}=
  2.04^{+0.37}_{-0.38}\times10^{13} (kT)^{1.65^{+0.11}_{-0.09}}$ for
  the distant clusters.  In comparison to their flux-limited sample
  (HIFLUGCS) and the sample with temperature profiles, the slope is in
  a good agreement within the errors but 
 it should be noted that the best-fit normalization is
  about 30\% smaller for the distant sample. This may be attributed
  to the significant redshift dependency of the critical density  
and will be worth further investigation in the light of the cluster formation redshift.

We also showed the $E_z M-T$ relation calculated for
$\Delta_c=500\Omega^{0.427}$ in Table 9.  
The result is
within a range consistent with the relation for $z>0.4$
\citep{Ettori_etal_2004}.

\item $M_{\rm gas}-T$ \\ 
The gas mass within $r_{500}$, $M_{\rm gas}$, is strongly
correlated to the temperature ($R_{XY}=0.61$). The fitting gives 
\begin{equation}
M_{\rm gas}~[h_{70}^{-5/2}{\rm M_{\odot}}]
=2.88^{+0.93}_{-0.64}\times10^{12} (kT)^{1.85^{+0.14}_{-0.15}},
\end{equation}
and $\chi^2/{\rm d.o.f}$ is 548.1/75.  Taking into account the current
statistical errors, we found a marginal steepening of the relation in
comparison to that predicted from the self-similar model, $M_{\rm
  gas}\propto T^{1.5}$ (see Fig.~\ref{fig16}g).  The measured slope is
also found to be consistent with that of the $M_{500}-T$ relation
derived above within their errors.

 For nearby clusters, \cite{Mohr_etal_1999} obtained the relation
  to be $M_{\rm gas}= (1.49\pm0.09)\times10^{14} (kT/6~{\rm
    keV})^{1.98\pm0.18}$ under $\Omega_0=1$ 
    and $H_0=50~{\rm km\,s^{-1}Mpc^{-1}}$,
  which is significantly steeper than the theoretically expected slope
  of 1.5.  On the other hand, \cite{Vikhlinin_etal_1999} reported
  based on the {\it ROSAT} PSPC data analysis of nearby regular
  clusters, a flatter relation in the form of $M_{\rm gas}\propto
  T^{1.71\pm0.13}$, utilizing a different method in determining the
  limiting radius (they defined the baryon overdensity radius of
  $R_{1000}$, corresponding to the dark matter overdensity of $\sim
  500$).

If we calculate the distant $M_{\rm gas}-T$ relation within $r_{500}$
using the same set of cosmological parameters, $M_{\rm gas}=
4.57^{+1.58}_{-1.17}\times10^{12} (kT)^{1.86^{+0.16}_{-0.16}}$.  The
relation is slightly flatter than that derived by
\cite{Mohr_etal_1999} but within a range consistent with the
  result of either \cite{Mohr_etal_1999} or
\cite{Vikhlinin_etal_1999} under the current measurement errors.

As shown above, although the relation within $r_{500}$ obtained for our
distant sample is consistent with the local relations, it is found to
be less steep than that found for $z>0.4$, $E_zM_{\rm gas}\propto
T^{2.37\pm0.17}$ \citep{Ettori_etal_2004}, which is not conflict with
the view that the lower redshift clusters contain more gas
for a fixed temperature, as pointed by \cite{Ettori_etal_2004}.

\item $f_{\rm gas}-T$ \\
We obtained a small correlation coefficient, $R_{XY}=-0.34$. 
The power-law fitting resulted in 
\begin{equation}
f_{\rm gas}~[h_{70}^{-3/2}] 
= 0.28^{+0.09}_{-0.06}(kT)^{-0.08^{+0.14}_{-0.14}},
\label{eq:fgas_t}
\end{equation}
and $\chi^2/{\rm d.o.f}=349.8/75$.  Because the resultant slope includes 0
within the error, we conclude that there is not a 
significant temperature dependence in the distant sample.  Thus we
show the mean gas-mass fraction of the sample, $\langle f_{\rm
gas}\rangle=0.20$, in Fig. \ref{fig16}h.
\end{itemize}

In the above analysis, we have not included the effect of the cluster
redshift. However, because a weak redshift dependence is seen in the
overdensity radius, which is $r_{500}\propto (1+z_{\rm obs})^{-0.6}$
(see \S\ref{subsubsec:rvir_fgas}), we checked how the parameter
correlations in equations \ref{eq:rvir_t}--\ref{eq:fgas_t} will be
changed if we divide the sample into the low-$z$ ($0.1<z\leq 0.3$) and
the high-$z$ ($0.3<z<0.82$) subsamples. As a result, there are
 no significant changes in the scaling relations compared to
  equations \ref{eq:rvir_t}--\ref{eq:fgas_t} except that the $M_{\rm
    gas}-T$ for the high-$z$ subsample resulted in a marginally
  steeper slope of $M_{\rm gas}=5.4^{+11.2}_{-4.0}\times10^{11}
  (kT)^{2.67^{+0.70}_{-0.58}}$ ($\chi^2/{\rm d.o.f}=170.2/30$).  In
  order to place firmer constraints on the scaling relations, we
  suggest that it is important to gather more observational data with
  higher sensitivities and also reexamine the assumption of the
  isothermal gas distribution and the formation redshift, $z_{\rm
    col}$ as already pointed out in \S\ref{subsubsec:rvir_fgas}. 

\subsection{Correlations with the core radius}\label{subsec:corr_rc}

\begin{figure}
\centering
\caption{Venn diagrams which illustrate the relation between the X-ray
  morphology and the optical morphology (a) and the X-ray core radius
  and the optical morphology (b) (see
    section~\ref{subsec:corr_rc} for definition of the cD cluster.)
  In the panel (b), we divided the sample into three subgroups: small
  core ($r_c < 0.1$ Mpc) single-$\beta$, double-$\beta$, and large
  core ($r_c >0.1$ Mpc) single-$\beta$ clusters. }
\label{fig17}
\end{figure}

\begin{figure*}
\centering
\caption{Relations of $\beta$ (a), the central electron density (b),
the cooling timescale (c), the temperature(d), and the overdensity radius 
(e) with the core radius. The meanings of the symbols are the same as
Fig. \ref{fig16}.  The error bars are $1\sigma$.  In the panels
(a)--(e) the best-fit power-laws obtained for 77 distant clusters are
shown with the solid lines. In the panels (b) and (c), the best-fit
power-laws for $\log r_c\leq -1$ and $\log r_c >-1$ are also shown
with the dashed lines and the dot-dash lines, respectively.  In the
panel (e) four dotted lines correspond to four different constant
values of $r_{500}/r_c$.}
\label{fig18}
\end{figure*}

\begin{itemize}

\item (Optical morphology)$-r_c$ \\ First we investigate the relation
to the optical morphology of the clusters.  Some clusters contain a
central dominant elliptical galaxy, i.e. a cD galaxy. We refer to such
clusters classified as Bautz-Morgan types I and I--II as ``cD
clusters''. We looked up the BM types of our sample clusters in the
NED database and showed the relation between X-ray morphology and the
BM type, and the core radius and the BM type in Fig. \ref{fig17}a
and \ref{fig17}b respectively. In Fig. \ref{fig17}b we divided the
clusters into three subgroups: the small core single-$\beta$, the
large core single-$\beta$ and the double-$\beta$ clusters. We find
that all the cD clusters are regular clusters and the clusters
with a small core (i.e. small core single-$\beta$ + double-$\beta$)
tend to contain a cD galaxy. However, not all the small core clusters
have cD galaxies, and the fraction having a cD galaxy is 36\%
(5 of 14). Therefore it is not a simple one-to-one
correspondence. In nearby clusters, the typical X-ray core radius of
cD galaxy is measured to be $\sim10$ kpc \citep{Ikebe_etal_1999} 
and is significantly smaller than 50 kpc. Thus it
is unlikely that the small core represents the potential distribution
of the cD galaxy itself, though some connection may be possible.  At
present, however, the data of the optical morphology is available for
only 47\% of the sample.  Thus in order to clarify the correlation
between the central galaxy and the formation of the small core, we
need to collect more optical data.  Though the above discussion was
based on the BM-type classification, we suggest that it is also
meaningful to take into account the existence of giant ellipticals at
the cluster center.

\item $\beta-r_c$\\
There is a weak trend of larger $\beta$ for larger $r_c$
($R_{XY}=0.62$). The fit yields
\begin{equation}
\beta = 0.73^{+0.07}_{-0.05}r_c^{0.11^{+0.03}_{-0.02}},
\end{equation}
and $\chi^2/{\rm d.o.f}$ is 711.1/75.  However the fit is not
statistically acceptable due to the huge $\chi^2$ value.  Regarding
this, we have to be careful about the parameter coupling in the
$\beta$ model fitting because the correlation seems to follow the
direction of the coupling (Fig. \ref{fig5}). This occurs noticeably
at $r_c\gtrsim 0.1$ Mpc. The correlation is strong for the large core
clusters with $r_c > 0.1$ Mpc ($R_{XY}=0.72$) while it is weak 
($R_{XY}=0.09$) for the small core clusters with $r_c<0.1$ Mpc. In
addition, all the clusters that exhibit extremely large core radii
($r_c \gtrsim 0.4$ Mpc) are irregular clusters and their surface
brightness distributions are highly inhomogeneous or bimodal (Fig. \ref{figc1}). 
Accordingly the current spherical $\beta$-model can
cause the tight $\beta-r_c$ coupling particularly for the irregular
systems.

\item $n_{e0}-r_c$ \\ As is clear from Fig. \ref{fig14}c, the
central electron density, $n_{e0}$, has a double-peaked distribution,
similar to $r_c$.  We see from Fig. \ref{fig18}b that there is a
strong correlation between $n_{e0}$ and $r_c$. The correlation
coefficient is $R_{XY}=-0.85$ for 77 clusters.  From the $\chi^2$
fitting, we obtain
\begin{equation}
n_{e0}~[h_{70}^{1/2}{\rm cm^{-3}}] 
= 0.89^{+0.30}_{-0.23}\times 10^{-3}r_c^{-1.29^{+0.10}_{-0.11}}.
\end{equation}
$\chi^2/{\rm d.o.f}$ is 1185/75.
We also find that the slope tends to be steeper for the small $r_c$
clusters: if we fit the data points for two different $r_c$ ranges
separately, we obtain
\begin{eqnarray}
n_{e0}~[h_{70}^{1/2}{\rm cm^{-3}}] &=&
0.13^{+0.36}_{-0.09}\times 10^{-3}r_c^{-1.87^{+0.41}_{-0.37}}
~{\rm for}~r_c \leq 0.1, \\
n_{e0}~[h_{70}^{1/2}{\rm cm^{-3}}] &=&
1.32^{+1.30}_{-0.83}\times 10^{-3}r_c^{-1.10^{+0.43}_{-0.66}}
~{\rm for}~r_c > 0.1.
\end{eqnarray}
$\chi^2/{\rm d.o.f}$ are 650.3/39 and 373.4/34, respectively. Thus the
gas distribution in clusters with $r_c\leq 0.1$ Mpc is concentrated
more than expected from the relation for clusters with $r_c > 0.1$
Mpc. This may suggest that the small core and the large core
components have different physical natures and/or they are at different
stages of evolution.

\item $t_{\rm cool}-r_c$ \\ According to Figs. \ref{fig12}e and
\ref{fig18}c, $t_{\rm cool}$ is significantly shorter than the age of
the Universe for the small core clusters and then the radiative
cooling is suggested to be important. We find a very tight correlation
between $t_{\rm cool}$ and $r_c$ ($R_{XY}=0.87$ for 77 clusters).  The
$t_{\rm cool}-r_c$ relations are derived to be
\begin{eqnarray}
t_{\rm cool}~{\rm [yr]} &=&
9.55^{+2.23}_{-2.07}\times 10^{10}r_c^{1.31^{+0.08}_{-0.09}}
~{\rm for~77~clusters}, \\ 
t_{\rm cool}~{\rm [yr]} &=& 
31.6^{+51.1}_{-20.4}\times 10^{10}r_c^{1.68^{+0.29}_{-0.32}}
~{\rm for}~r_c \leq 0.1, \\
t_{\rm cool}~{\rm [yr]} &=& 
3.63^{+1.21}_{-0.91}\times 10^{10}r_c^{0.70^{+0.18}_{-0.18}}
~{\rm for}~r_c > 0.1.
\end{eqnarray}
$\chi^2/{\rm d.o.f}$ are 690.0/75, 402.4/39 and 175.9/34,
respectively. Thus one possible interpretation of the small core may
be that the small core radius does not reflect the shape of the
gravitational potential but that it reflects the cooling radius inside
which the X-ray emission is enhanced. However we consider this is
unlikely for the following reasons. If the small core radius reflects
the cooling radius, it should evolve with time. However we do not find
a strong redshift dependence in the core radius. Moreover we estimated
the cooling radius at which the cooling time of the gas becomes equal
to $t_{\rm age}$ to find it is larger than 50 kpc for most of the
small core clusters.

\item $T-r_c$ \\ Furthermore, as shown in Fig. \ref{fig18}d, we find
  that the temperature does not show a clear core radius dependence.
  We obtained a moderate correlation coefficient, $R_{XY}=0.45$.  The
  best-fit relation is derived as
\begin{equation}
kT =13.7^{+2.6}_{-2.5}r_c^{0.30^{+0.05}_{-0.07}}, 
\end{equation}
however, the power-law fit was quite poor ($\chi^2/{\rm
  d.o.f}=2228/75$). Since the emission-weighted temperature reflects
the temperature of the cluster core region, the apparent lack of any
strong $T-r_c$ correlation suggests that the temperature gradient in
the cooling region is not very large. This is consistent with the
results from the {\it XMM-Newton} observations, which revealed that
the temperature gradient of nearby cooling-flow clusters is smaller
than that expected from the standard cooling flow model
\cite[e.g.][]{Tamura_etal_2001}. \cite{Peterson_etal_2003} noted that
there is a significant deficit of emission with temperature lower than
$T_0/3$ ($T_0$ is the ambient temperature) in the RGS spectra. Given
that the temperature profile obeys $T(r)\propto r^{0.2}$, the second
term in Equation~\ref{eq:hydrostatic} is negligible compared to the
first term at $r\gtrsim0.3r_c$ for $\beta=2/3$. Thus except for the
central $r\lesssim 0.3 r_c$ region, the cluster mass profile can be
approximated with the assumption of a constant temperature, suggesting
the gas density profile obtained from the current isothermal
$\beta$-model analysis reflects the underlying cluster potential
distribution. We further investigate the relation between $r_{500}$
and the core radius below.

\item $r_{500}-r_c$ \\ Since the overdensity radius is
  determined almost independently from $r_c$ and approximately
  $r_{500}\propto T^{1/2}$, the above results indicate that
  correlation between $r_{500}$ and $r_c$ is much weaker than the
  expectations of the self-similar model.  It is clear from
  Fig. \ref{fig18}e that the distribution of the data points are
  inconsistent with the curves of $r_{\rm 500}/r_c={\rm
    constant}$. The observed range of $r_{\rm 500}/r_c$ is about
    2--40.  The correlation coefficient is $R_{XY}=0.58$ for 77
  clusters. Then the $\chi^2$ fitting gives
\begin{equation}
r_{500}~[h_{70}^{-1}{\rm Mpc}] = 
1.51^{+0.16}_{-0.17}r_c^{0.15^{+0.03}_{-0.04}}
~{\rm for~77~clusters}, \label{eq:rvir_rc_all}
\end{equation}
and $\chi^2/{\rm d.o.f.}=954.5/75$. We notice that the observed
distribution on the $r_{500}-r_c$ plane is much flatter than the
curves for constant $r_{500}/r_c$ values or it is rather
concentrated around the two peak values of $r_c$.  The departure from
the self-similar relation is more prominent for small core clusters
with $r_c<0.1$ Mpc. The correlation coefficients are $R_{XY}=0.21$ and
0.36 for $r_c\le0.1$ and $r_c>0.1$, respectively. Thus for the large
core clusters, the best-fit power-law relation is
\begin{equation}
r_{500}~[h_{70}^{-1}{\rm Mpc}] =
2.06^{+0.42}_{-0.33}r_c^{0.37^{+0.12}_{-0.11}}
~{\rm for}~r_c > 0.1, \label{eq:rvir_rc_large}
\end{equation}
and $\chi^2/{\rm d.o.f.}$ is 108.0/34.  If we further restrict $r_c$
to a very narrow range of 0.1--0.2 Mpc, we find a steeper slope of
$r_{500}=4.03^{+3.22}_{-1.52}r_c^{0.71^{+0.30}_{-0.26}}$
($\chi^2/{\rm d.o.f}=21.2/18$). Thus those twenty clusters may satisfy
the self-similar condition, $r_{500}\propto r_c$. However, we
suggest from equations \ref{eq:rvir_rc_all}--\ref{eq:rvir_rc_large}
that it is difficult to explain the formation of the cores,
particularly for the small core clusters, by the standard picture of
the self-similar model. 
\end{itemize}

\subsection{Implications on the origin of two core scales}

From the above discussion, it seems difficult to explain the small
core size that we discovered in the histogram either by the potential
structure of the cD galaxy or the cooling radius. As long as we rely
on the hydrostatic assumption and the $\beta$ model, the X-ray surface
brightness distributions are likely to represent the gravitational
potential structures of the clusters.  If this is the case, the
double-$\beta$ nature of the X-ray emission profile reflects the shape
of the gravitational potential of the dark matter, which is likely to
have two preferable scales of $\sim 50$ kpc and $\sim 200$ kpc. 

In an effort to constrain the physical status of ICM in the dark matter
potential, comparing the high-resolution X-ray observations to
gravitational lensing observations will provide another powerful test
\citep[e.g.][]{Hattori_etal_1997}. Thanks to the improvement of
spatial resolution achieved by {\it Chandra}, now measurements on the
cluster mass profile down to $\lesssim 5$ kpc scale are possible at
such high redshifts ($z\sim 0.3$) and several authors have measured
the dark matter distribution in the lensing clusters under the
hydrostatic hypothesis \citep[e.g.][]{Arabadjis_etal_2002,
  Xue_Wu_2002, Ota_etal_2004}.  For example, \cite{Ota_etal_2004} 
  showed from the high-resolution {\it Chandra} data of
\object{CL0024+17} ($z=0.395$) and the comparison with the detailed
lens modeling by \cite{Tyson_etal_1998} that the cluster density
profile is well reproduced by the double-$\beta$ model and the inner
core also reflects the underlying dark matter potential. They also
noted that the core structure may be related to the past merging event
as inferred from the optical observations \citep{Czoske_etal_2001,
  Czoske_etal_2002}.  
  
 Recently \cite{Hayakawa_etal_2004} estimated the dark matter
  distribution in a nearby non-cD, regular cluster, Abell 1060, from
  the {\it Chandra} data analysis, without explicitly using the double
  $\beta$-model, and found a central mass concentration at $r<
  50$~kpc.  Their result also supports the idea that dark matter may
  preferentially be accumulated within a radius of $\sim 50$~kpc.  On
the other hand, \cite{Ettori_etal_2004} suggested that no significant
double structure is seen in the high redshift sample.  Thus
considering from the above, the double-$\beta$ nature of the ICM
discovered in the present sample may be much related to the history of
the past merging and the relaxation process. 
The small core component might be attributed
to the presence of dark matter subhalos due to the cluster mergers or
the internal structures in clusters \citep[e.g.][]{Fujita_etal_2002}.

Furthermore, since there is clearly a tight coupling between the core
radius and the radiative cooling time as shown in section
\ref{subsec:corr_rc}, the detailed treatment of the thermal evolution
of the ICM will also be important. A number of numerical simulations
including non-gravitational effects such as radiative cooling and
galaxy feedback have been carried out and thus provide a clue 
to the underlying physics in the cluster core regions. However, the
authors pointed out difficulties in regulating the central over-cooling
and produce a constant-density core \citep[e.g.][]{Pearce_etal_2000}.
\cite{Masai_Kitayama_2004} recently proposed a quasi-hydrostatic
model, which predicts a characteristic temperature profile with an
asymptotic temperature for the central region being $\sim1/3$ of the
non-cooling outer region, as observed in nearby ``cooling flow''
clusters. Thus detailed comparison of the X-ray data with their model
regarding the temperature and density profiles will be important to
understand the evolution of the ICM structure.  We need further
investigations to put a stronger constraint on the origin of the
double nature of the cluster structures, which is however beyond the
scope of the present paper and will be discussed in a separate paper.

\section{Summary}\label{sec:summary}
We have analyzed the {\it ROSAT} HRI and the {\it ASCA} GIS/SIS data
of 79 clusters of galaxies at redshifts of 0.1 -- 0.82 in a uniform
manner.  We determined the X-ray surface brightness profile from the
{\it ROSAT} HRI data utilizing the $\beta$-model and the average
temperature and the luminosity from the {\it ASCA} data.  We found
that the clusters can be divided into two subgroups, regular and
irregular clusters, from analysis to determine the centroid position
of the X-ray image. We then performed a statistical study of the
X-ray parameters and investigated the trends for redshift
evolution and the scaling relations against temperature and core
radius. The major results are summarized as follows.

\begin{enumerate}

\item We did not find significant redshift evolution in the X-ray
  parameters of clusters compared to the nearby clusters: the
  temperature $kT$, the core radius $r_c$, $\beta$, and the central
  electron density $n_{e0}$ at $z\lesssim 0.5$.

\item Among the X-ray parameters, the core radius shows the largest
  cluster-to-cluster variation. The core-radius distribution shows two
  distinct peaks at 50 kpc and 200 kpc.  For 20 \% of the regular
  clusters, inclusion of a second $\beta$-model component
  significantly improved the $\chi^2$ values of the surface-brightness
  fitting.  We find that the two core radii of the double
  $\beta$-model are distributed in relatively narrow ranges consistent
  with the two peaks of the single-$\beta$ clusters. There is no 
  significant evolution in the fraction of double-$\beta$ clusters
  within the observed redshift range.

\item We investigated the correlations between the temperature and the
  cluster parameters including the spectral and the $\beta$-model
  parameters, the cluster mass, 
  the gas mass and the gas-mass fraction etc.  
For the $M_{500}-T$ relation, we found that 
  the power-law slope of $1.68^{+0.10}_{-0.11}$ 
  is marginally steeper than that expected from the self-similar model but in a good agreement with the results for the nearby clusters, while the normalization factor is about 30\% smaller for the current distant sample compared to the nearby sample. 
  We obtained the $M_{\rm gas}-T$ relation to be $M_{\rm gas}\propto
    (kT)^{1.86^{+0.16}_{-0.16}}$ for the overdensity of
    $\Delta_c=500$ and found a marginal steepening of the relation
  in comparison to the self-similar model 
  under the current statistics.  The $f_{\rm gas}-T$ relation is found
  to be consistent with having no correlation with the temperature.

\item We studied the parameter correlations against the core
  radius. We found that only 36\% of the small core single-$\beta$ and
  the double-$\beta$ clusters are cD clusters and thus it seems
  difficult to explain the presence of the small core by the cD
  potential itself although there may be some causal link.  There
  are strong $n_{e0}-r_c$ and $t_{\rm cool}-r_c$ correlations and the
  slopes tend to become steeper for $r_c \lesssim 0.1$ Mpc.  On the
  other hand the fact that there is not a clear $T-r_c$ correlation
  suggests that the temperature gradient is not large even in clusters
  with short cooling timescales, which is consistent with the {\it
    XMM-Newton} and {\it Chandra} observations of the nearby clusters.
   Thus as long as we rely on the hydrostatic condition and the
    $\beta$-model, our result indicates that the dark matter
    distribution is likely to show two preferable scales of 50~kpc and
    200~kpc.

\item We showed that the $r_{500}-r_c$ relation derived from the X-ray
  analysis does not agree with the expectations of the self-similar
  model, suggesting that the assumption of self-similarity is not valid in 
  describing the density profile of the ICM, particularly for clusters
  with small core radius.

\item  We obtained the average gas-mass fraction within $r_{500}$
  to be $\langle f_{\rm gas}\rangle = (0.20\pm 0.07)~h_{70}^{-3/2}$
  for the distant sample.  The calibrational error is estimated to be
  about 25\%.  The current estimation is based on some simplified
  assumptions (for example, the isothermal gas distribution), 
which will be refined in future studies. 
  
\end{enumerate}

\begin{acknowledgements}
We are grateful to S. Sasaki, T. Kitayama and K. Masai for their
helpful comments and discussions. N.O. is supported by a Research
Fellowship for Young Scientists from the JSPS. This research has made
use of the {\it ROSAT} HRI data obtained through the {\it ROSAT} Data
Archive of MPE at Garching, Germany and the {\it ASCA} data through
the DARTS Online Service provided by ISAS (ISAS/JAXA) and the
HEASARC Online Service provided by NASA/Goddard Space Flight Center.
We also thank the anonymous referee for helpful comments. 
\end{acknowledgements}

\appendix

\section{Mass profile of double-$\beta$ clusters}\label{appendix:doubleb}

Suppose that the density distribution of intracluster gas is
characterized by superposition of the two $\beta$-model gas profiles.
\begin{equation}
\rho_{gas}(r) = \sum_{i=1}^{2} \rho_{gas,i}(0)
\left[1+\left(\frac{r}{r_i}\right)^2\right]^{-3\beta_i/2}.
\label{eq:gasdensity_db}
\end{equation}
Then the X-ray surface brightness distribution is given by integrating the 
X-ray emissivity along the line of sight, 
\begin{eqnarray}
S(r) &=& \int dl \frac{\epsilon_{\rm ff}}{4\pi D_L^2} 
\sim\int dl (\sum_{i=1}^{2}{n_i(r)}^2\Lambda(T,Z))
\frac{1}{4\pi D_L^2}, \nonumber \\ 
& = &
\sum_{i=1}^{2}S_i\left[1+\left(\frac{r}{r_i}\right)^2\right]^{-3\beta_i+1/2}.
\end{eqnarray}
We refer to the second line of the above equation as the double
$\beta$-model and utilized in \S\ref{subsec:wbetafit}.  We also obtain
the thermal pressure of the gas:
\begin{equation}
P(r) = \sum_{i=1}^{2} n_i(r)kT.
\end{equation}
Then the total cluster mass is estimated from the hydrostatic equation to be
\begin{eqnarray}
M(r) &=& -\frac{kTr^2}{\mu m_p G}\frac{\partial \ln{n(r)}}{\partial r}, \nonumber\\ 
& = & \frac{3kT r^3}{\mu m_pG}\frac{\sum_{i}\beta_in_{0,i}r_i^{-2}
[1+(\frac{r}{r_i})^2]^{-3\beta_i/2-1}}{\sum_{i}n_i(r)}. \label{eq:mvir_db}
\end{eqnarray}
The average density profile of the total cluster mass is then 
\begin{equation}
\bar{\rho}(r) = \frac{M(r)}{\frac{4}{3}\pi r^3} = \frac{9kT}{4\pi\mu m_p G}
\frac{\sum_{i}\beta_in_{0,i}r_i^{-2}
[1+(\frac{r}{r_i})^2]^{-3\beta_i/2-1}}{\sum_{i}n_i(r)}.\label{eq:average_rho_db}
\end{equation}
We define the central cluster mass density as
\begin{equation}
\rho_0 \equiv \bar{\rho}(r)|_{r\rightarrow0} = \frac{9kT}{4\pi \mu
m_pG}\frac{\sum_{i}\beta_in_{0,i}r_i^{-2}}{\sum_{i}n_{0,i}}. \label{eq:rho0}
\end{equation}

\section{Systematic error of gas-mass fraction}\label{appendix:syserr}

The gas-mass fraction is determined from the parameters, $T$, $\beta$,
$r_c$, and $n_{e0}$, and $n_{e0}$ is further determined from the central
surface brightness $S_{\rm p0}$, $T$, $r_c$, and $\beta$.  Among those
parameters, a possible systematic effect for $r_c$ was investigated in
section \ref{subsec:sbetafit}, and was found to be less than the pixel
size of the X-ray image.  $\beta$ is strongly coupled to $r_c$ and its
systematic error is determined by the systematic error of $r_c$. On
the other hand, $n_{e0}$ requires the absolute calibration of the
X-ray telescope/detector effective area, which usually contains large
systematic errors.  The temperature, $T$, requires a calibration of
the effective area as a function of X-ray energy.  Thus we will
investigate the systematic errors in $n_{e0}$ and $T$ below.

Systematic errors in the electron density mainly come from the
calibration of the {\it ROSAT} HRI.  Because the gain of the HRI
continuously decreased from the launch to the end of the mission, the
conversion factor from the photon energy to pulse height is time
dependent.  Thus the errors in the gain determination cause
significant error in the absolute flux.  The possible range of the
gain variations over the lifetime of {\it ROSAT} was measured by
\cite{Prestwitch_etal_1998}.  Thus we tried the response matrices for
the two extreme gain values in calculating Equation
\ref{eq:beta_sp0_to_ne0} to estimate the maximum systematic errors.
Taking into account that the result also depends on the cluster
emission spectrum, the effect is $\sim 20$\% in the worst case.  The
calibration of the effective area of the X-ray telescope and the HRI
system is also reported by the {\it ROSAT} Science Data Center, from
which we estimate that the systematic errors in $n_{e0}$ is $\sim$
10\%.

Systematic errors in the temperature come from the calibration of {\it
ASCA}.  The response functions of the {\it ASCA} XRT/GIS and XRT/SIS
are well calibrated for point sources. However, for the extended
sources, there still are significant systematic errors.  In
particular we found that the derived temperature is dependent on the
spectrum integration region on the detector if the integration radius
is too small. In the spectral analysis, we determined the radius so
that the dependence becomes insignificant.  However, we still consider
there is some systematic effect related to this problem and estimate
that it will affect $f_{\rm gas}$ by about 5\% (notice $T$ affects
both $M_{500}$ and $M_{\rm gas}$).

In total, the systematic error due to the instrument calibrations is
estimated to be $\sim 25$\%.

\section{Individual clusters}
\begin{figure}
\centering
\caption{{\it ROSAT} HRI images, radial surface brightness profiles
  and the {\it ASCA} SIS and GIS spectra of 79 clusters. The HRI
  images are smoothed by a Gaussian filter with
  $\sigma=7\arcsec.5-15\arcsec$ and the contours whose levels
  correspond to $n (=3, 5, 9, 15, 31, 63)$ times the $1\sigma$
  background level are overlaid. The backgrounds are not
  subtracted. The X-ray centroids that were determined and used to
  derive the radial profiles in the image analysis are marked with the
  crosses. In the central panels, the crosses denote the observed
  radial profile of the HRI, and the step functions show the best-fit
  $\beta$ models. The best-fit background levels are shown with the
  dashed lines. For the nine double $\beta$ clusters, the results of
  the double $\beta$ model fitting are shown instead of the single
  $\beta$ model and the inner and the outer components are also shown
  with the dash-dot and the dotted lines, respectively.  The X-ray
  significance radius, $r_x$, and the overdensity radius,
    $r_{500}$ are shown with the vertical dashed, and dotted lines
    respectively. In the right panels, the {\it ASCA} spectra fitted
  with the Raymond-Smith model are shown, where the crosses and
  crosses with circles denote the spectra obtained with the GIS and
  the SIS respectively, and the stepped lines show the best-fit models
  convolved with the telescope and detector responses.  The fitting
  residuals are also shown in the panels. \label{figc1}}
\end{figure}

\end{document}